\documentclass[a4paper,UKenglish]{lipics-v2021}

\hideLIPIcs  

\title{Structure-Guided Automated Reasoning}
\titlerunning{Structure-Guided Automated Reasoning}

\author{Max Bannach}{European Space Agency, Advanced Concepts Team, Noordwijk, The Netherlands}{max.bannach@esa.int}{https://orcid.org/0000-0002-6475-5512}{}
\author{Markus Hecher}{Massachusetts Institute of Technology, Computer Science and Artificial Intelligence Lab, USA}{hecher@mit.edu}{https://orcid.org/0000-0003-0131-6771}{}

\authorrunning{M. Bannach and M. Hecher}
\Copyright{Max Bannach and Markus Hecher} 

\category{} 

\relatedversion{} 

\nolinenumbers 

\EventEditors{John Q. Open and Joan R. Access}
\EventNoEds{2}
\EventLongTitle{42nd Conference on Very Important Topics (CVIT 2016)}
\EventShortTitle{CVIT 2016}
\EventAcronym{CVIT}
\EventYear{2016}
\EventDate{December 24--27, 2016}
\EventLocation{Little Whinging, United Kingdom}
\EventLogo{}
\SeriesVolume{42}
\ArticleNo{23}

\usepackage{amssymb,mathtools}
\usepackage{tikz}
\usetikzlibrary{graphs, arrows.meta, calc, backgrounds}
\usetikzlibrary{intersections}
\usepackage{booktabs}
\usepackage{pifont}

\usepackage[nomove]{tcs-automove}

\DeclareMathOperator{\tower}{\mathrm{tower}\kern-0.2ex^*\kern-0.2ex}
\DeclareMathOperator{\toweronly}{\mathrm{tower}}
\DeclareMathOperator{\qa}{\mathrm{qa}}

\newcommand\Class[1]{%
  \mathchoice%
  {\text{\normalfont\small$\mathrm{#1}$}}%
  {\text{\normalfont\small$\mathrm{#1}$}}%
  {\text{\normalfont$\mathrm{#1}$}}%
  {\text{\normalfont$\mathrm{#1}$}}%
}
\newcommand{\Lang}[1]{\text{\normalfont\textsc{#1}}}

\def\phi{\varphi}
\DeclareMathOperator{\edges}{atoms}

\DeclareMathOperator{\suchthat}{\centerdot}
\def\struc{\textsc{struc}}
\def\structure{\mathcal{S}}
\def\graphstructure{\mathcal{G}}
\def\lang{\mathcal{L}}
\DeclareMathOperator{\width}{\mathrm{width}}
\DeclareMathOperator{\tw}{\mathrm{tw}}

\DeclareMathOperator{\qr}{\mathrm{qr}}
\DeclareMathOperator{\bs}{\mathrm{bs}}
\DeclareMathOperator{\bag}{\chi}
\DeclareMathOperator{\baglabel}{\lambda}

\DeclareMathOperator{\vars}{\mathrm{vars}}
\DeclareMathOperator{\compat}{\mathrm{compat}}
\newcommand\var{\vars}
\DeclareMathOperator{\lits}{\mathrm{lits}}
\DeclareMathOperator{\clauses}{\mathrm{clauses}}
\DeclareMathOperator{\termsx}{\mathrm{terms}}
\DeclareMathOperator{\soft}{\mathrm{soft}}
\DeclareMathOperator{\hard}{\mathrm{hard}}

\DeclareMathOperator{\assignment}{\sqsubseteq}

\DeclareMathOperator{\card}{\mathrm{card}}

\def\comment#1{\text{\color{gray}\it#1}}%
\DeclareMathOperator{\children}{\mathrm{children}}
\DeclareMathOperator{\rootOf}{\mathrm{root}}
\DeclareMathOperator{\parent}{parent}
\DeclareMathOperator{\atoms}{atoms}
\def\sat{\mathit{sat}}

\definecolor{jade}{rgb}{0.0, 0.66, 0.42}
\definecolor{cerise}{HTML}{CE4760}
\colorlet{fg}{jade!75!black}
\colorlet{bg}{cerise!75!black}

\tikzset{
  picograph/.style = {
    nodes = {draw, circle, semithick, inner sep=1pt},
    semithick,
    >={[round, scale=0.5]Stealth}
  }  
}

\def\picocycle{%
  \tikz[baseline={([yshift=0.5]a)}, picograph, scale = 0.25]{
    \node (a) at (0,0) {};
    \node (b) at (1,0) {};
    \node (c) at (1,1) {};
    \node (d) at (0,1) {};
    \graph[use existing nodes]{ a -> b -> c -> d -> a };
  }%
}

\def\picoclique{%
  \tikz[baseline={([yshift=0.5]a)}, picograph, scale = 0.25]{
    \node (a) at (0,0) {};
    \node (b) at (1,0) {};
    \node (c) at (1,1) {};
    \node (d) at (0,1) {};
    \graph[use existing nodes]{ a -- b -- c -- d -- a; a -- c; b -- d; };
  }%
}

\def\picocycleundir{%
  \tikz[baseline={([yshift=0.5]a)}, picograph, scale = 0.25]{
    \node (a) at (0,0) {};
    \node (b) at (1,0) {};
    \node (c) at (1,1) {};
    \node (d) at (0,1) {};
    \graph[use existing nodes, edges = {bend left=15}]{ a -> b -> c -> d -> a };
    \graph[use existing nodes, edges = {bend right=15}]{ a <- b <- c <- d <- a };
  }%
}

\def\picoevencycle{%
  \tikz[baseline={([yshift=0.5]a)}, picograph, scale = 0.25]{
    \node (a) at (0,0) {};
    \node (b) at (1,0) {};
    \node (c) at (1,1) {};
    \node (d) at (0,1) {};
    \graph[use existing nodes]{ a -- b -- c -- d -- a };
  }%
}

\newtheorem{fact}{Fact}

\begin{document}

\maketitle

\begin{abstract}
  Algorithmic meta-theorems state that problems definable in a fixed
  logic can be solved efficiently on structures with
  certain properties. An example is Courcelle's Theorem, which states
  that all problems expressible in monadic second-order logic can be
  solved efficiently on structures of small treewidth. Such theorems
  are usually proven by algorithms for the model-checking
  problem of the logic, which is often complex and rarely leads
  to highly efficient solutions. Alternatively, we can solve the
  model-checking problem by grounding the given logic to propositional
  logic, for which dedicated solvers are available. Such encodings
  will, however, usually not preserve the input's treewidth.

  This paper investigates whether all problems definable in monadic
  second-order logic can efficiently be encoded into \Lang{sat} such
  that the input's treewidth bounds the treewidth of the resulting
  formula. We answer this in the affirmative and, hence, provide an
  alternative proof of Courcelle's Theorem. Our technique can
  naturally be extended: There are treewidth-aware reductions from the
  optimization version of Courcelle's Theorem to \Lang{maxsat} and
  from the counting version of the theorem to \Lang{\#sat}.  By using
  encodings to \Lang{sat}, we obtain, ignoring polynomial factors, the
  same running time for the model-checking problem as we would with
  dedicated algorithms. Another immediate consequence is a
  treewidth-preserving reduction from the model-checking problem of
  monadic second-order logic to integer linear programming
  (\Lang{ilp}). We complement our upper bounds with new lower
  bounds based on $\Class{ETH}$; and we show that the block size of
  the input's formula and the treewidth of the input's structure are
  tightly linked.
  
  Finally, we present various side results needed to prove the main
  theorems: A treewidth-preserving cardinality constraints, treewidth-preserving
  encodings from \Lang{cnf}s into \Lang{dnf}s, and a treewidth-aware
  quantifier elimination scheme for \Lang{qbf} implying a
  treewidth-preserving reduction from \Lang{qsat} to \Lang{sat}. We also present a reduction from
  projected model counting to $\Lang{\#sat}$ that increases the
  treewidth by at most a factor of $2^{k+3.59}$, yielding a algorithm
  for projected model counting that beats the currently best running
  time of $2^{2^{k+4}}\cdot\mathrm{poly}(|\psi|)$.
\end{abstract}

\clearpage
%
%
\section{Introduction}
\label{section:introduction}

Many tools from the automated reasoning quiver can be implemented
efficiently if a graphical representation of the given formula with
good structural properties is given. The textbook example is the
\emph{satisfiability problem} (\Lang{sat}), which can be solved in
time $O\big(2^k\mathrm{poly}(|\psi|)\big)$ on formulas~$\psi$ whose
\emph{primal graph} $G_\psi$ has \emph{treewidth} $k$. (The primal
graph contains a vertex for every variable of the formula and connects
them if they appear together in a
clause. Its treewidth intuitively measures how close it is to being a
tree.) The result extends to the \emph{maximum satisfiability problem}
(\Lang{maxsat}), in which the clauses of the formula have weights and
the goal is to minimize the weights of 
falsified clauses, and to the \emph{model counting problem}
(\Lang{\#sat}), in which the goal is to compute the \emph{number} of
satisfying assignments. In this article, we
will use the notation $\toweronly(h,t)$ to describe a tower of twos of
height $h$ with $t$ at the top, and $\tower(h,t)$ as shorthand to hide
polynomial factors, e.g., $O\big(2^k\mathrm{poly}(|\psi|)\big)=\tower(1,k)$:

\begin{fact}
  [folklore, see for
    instance~\cite{AlekhnovichR02,BacchusDP03,BannachST22,Darwiche01,KorhonenJ21,SaetherTV15,SamerS10}]
  \label{fact:semiringtw}
  One can solve \Lang{sat}, \Lang{maxsat}, and
  \Lang{\#sat} in time $\tower(1,k)$ if a
  width-$k$ tree decomposition is given.
\end{fact}

It is worth to take some time to inspect the details of
Fact~\ref{fact:semiringtw}. The hidden polynomial factor is not the
subject of this paper (as indicated by the notation), but can be made
as small as $O(|\phi|)$~\cite{CapelliM18,LampisMM18}. Our focus will
be the value on top of the tower, which in Fact~\ref{fact:semiringtw}
is simply ``$k$''. Under the
\emph{exponential-time hypothesis} ($\Class{ETH}$), this is best possible.

The natural extension of the satisfiability problem to higher logic
is the validity problem of fully \emph{quantified Boolean formulas}
(\Lang{qsat}). While it is well-known that \Lang{qsat} is
fixed-parameter tractable (i.e., it is in $\Class{FPT}$) with respect to treewidth~\cite{Chen04}, the
dependencies on the treewidth is less sharp than in
Fact~\ref{fact:semiringtw}. The height of the tower depends on the
\emph{quantifier alternation} $\qa(\psi)$ of the formula, while the top value has
the form $O(k+\log k+\log\log k+\dots)$ due to the management of nested
tables in the involved dynamic program.
\begin{fact}[\cite{Chen04,CapelliM19}]
  \label{fact:qsat}
  One can solve \Lang{qsat} in time
  $\tower\!\big(\qa(\psi)+1,O(k)\big)$ if a width-$k$ tree decomposition is given.
\end{fact}

In contrast to Fact~\ref{fact:semiringtw}, there is a big-oh
on top of the tower in Fact~\ref{fact:qsat}. The higher order version
of the model counting problem is the \emph{projected model
  counting problem} (\Lang{pmc}), in which we need to count the number of models that are
not identical on a given set of variables.

\begin{fact}[\cite{FichteHMTW23}]
  \label{fact:pmc}
  One can solve \Lang{pmc} in time
  $\tower\!\big(2,k+4\big)$ if a width-$k$ tree decomposition is
  given.
\end{fact}

The fine art of automated reasoning is \emph{descriptive complexity},
which studies the complexity of problems in terms of the complexity of
a description of these problems; independent of any abstract machine
model~\cite{Immerman99,Kreutzer11}. A prominent example is
\emph{Courcelle's Theorem} that states that the problems that can be
expressed in \emph{monadic second-order logic} can be solved
efficiently on instances of bounded
treewidth~\cite{Courcelle90}. Differently phrased, the theorem states
that the \emph{model-checking problem} for \Lang{mso} logic
(\Lang{mc(mso)}) is fixed-parameter tractable (the parameter is the
size of the formula and the  treewidth of the structure):
\begin{fact}[\cite{Courcelle90}]
  \label{fact:mso}
  One can solve \Lang{mc(mso)} in time
  $\tower\!\big(\qa(\phi)+1, O(k+|\phi|) \big)$ if a width-$k$ tree decomposition is given.
\end{fact}

For instance, the 3-coloring problem (Can we color the vertices of a graph with three colors such that
adjacent vertices obtain different colors?) can be described by the sentence:
\begin{align*}
  &\phi_{\mathrm{3col}} =
  \exists R\exists G\exists B\,\forall x\forall y\suchthat\, ( Rx \vee Gx \vee Bx )\\
  &\quad\wedge
  Exy \rightarrow \neg( (Rx\wedge Ry)\vee(Gx\wedge Gy)\vee(Bx\wedge By) ).
\end{align*}
The sentence can be read aloud as: There are three colors red, blue,
and green ($\exists R\exists G\exists B$) such that for all vertices
$x$ and $y$ ($\forall x\forall y$) we have that (\emph{i}) each vertex
has at least one color ($Rx \vee Gx \vee Bx$), and (\emph{ii}), if $x$
and $y$ are connected by an edge ($Exy$) then they do not have the
same color ($\neg( (Rx\wedge Ry)\vee(Gx\wedge Gy)\vee(Bx\wedge By) )$).
The model-checking problem \Lang{mc(mso)}
obtains as input a relational structure $\structure$ (say a graph like
\picoevencycle\ or \picoclique) and an \Lang{mso} sentence $\phi$
(as the one from above) and asks whether $\structure$ is a model of~$\phi$,
denoted by $\structure\models\phi$. In our example we have
$\picoevencycle\models\phi_{\mathrm{3col}}$ and $\picoclique\not\models\phi_{\mathrm{3col}}$. 
Using Fact~\ref{fact:mso}, we can conclude from $\phi_{\mathrm{3col}}$ that the
3-coloring problem parameterized by the treewidth lies in $\Class{FPT}$.

Instead of utilizing Fact~\ref{fact:mso}, another reasonable approach
is to \emph{ground} the \Lang{mso} sentence to a propositional formula
and to then apply Fact~\ref{fact:semiringtw}. Formally, this
means to reduce the model checking problem \Lang{mc(mso)} to \Lang{sat}, i.e., given a
relational structure $\structure$ and an \Lang{mso} sentence $\phi$,
we need to produce, in polynomial time, a propositional formula $\psi$
such that $\structure\models\phi$ iff $\psi\in\Lang{sat}$.  The
na\"ive way of doing so is by generating an indicator variable $X_u$
for every set variable $X$ and every element $u$ in the universe of
$\structure$. Then we replace every first-order $\exists$-quantifier by a ``big-or'' and $\forall$-quantifier by a
``big-and'':
\[
  \color{gray}
  {\color{black}\psi_{\mathrm{3col}} =}
  \overbrace{\color{black}\bigwedge_{u\in V(G)}\bigwedge_{v\in V(G)}}^{\forall x\forall y}
  \overbrace{\color{black}\vphantom{\bigwedge_{v\in V(G)}}%
    (\tikz[remember picture,baseline={(Rv1.base)}]\node[inner sep=1pt](Rv1){$R_u$};%
    \vee%
    \tikz[remember picture,baseline={(Gv1.base)}]\node[inner sep=1pt](Gv1){$G_u$};%
    \vee%
    \tikz[remember picture,baseline={(Bv1.base)}]\node[inner sep=1pt](Bv1){$B_u$};)%
  }^{Rx\,\vee\, Gx\,\vee\, Bx}
  {\color{black}\wedge}
  \overbrace{\color{black}\mathclap{\bigwedge_{\{u,v\}\in E(G)}}}^{Exy\rightarrow}
  {\color{black}
    \neg( (\tikz[remember picture,baseline={(Rv2.base)}]\node[inner sep=1pt](Rv2){$R_u$};%
    \wedge
    \tikz[remember picture,baseline={(Rw.base)}]\node[inner sep=1pt](Rw){$R_v$};
    )\vee
    (
    \tikz[remember picture,baseline={(Gv2.base)}]\node[inner sep=1pt](Gv2){$G_u$};
    \wedge
    \tikz[remember picture,baseline={(Gw.base)}]\node[inner sep=1pt](Gw){$G_v$};
    )
    \vee
    (
    \tikz[remember picture,baseline={(Bv2.base)}]\node[inner sep=1pt](Bv2){$B_u$};
    \wedge
    \tikz[remember picture,baseline={(Bw.base)}]\node[inner sep=1pt](Bw){$B_v$};
    ))
    .}
\]
\begin{tikzpicture}[remember picture, overlay]
  \node at (0,0) {};
  \node[color=gray] (text) at (9,0) {\it propositional variables};
  \draw[color=gray, semithick, rounded corners] (text) -| (Rv1);
  \draw[color=gray, semithick, rounded corners] (text) -| (Gv1);
  \draw[color=gray, semithick, rounded corners] (text) -| (Bv1);
  \draw[color=gray, semithick, rounded corners] (text) -| (Bv2);
  \draw[color=gray, semithick, rounded corners] (text) -| (Gw);
  \draw[color=gray, semithick, rounded corners] (text) -| (Bw);

  \draw[name path=line, color=gray, semithick] (Rv2.south) -- (Rv2.south |- text.north);
  \draw[name path=line, color=gray, semithick] (Gv2.south) -- (Gv2.south |- text.north);
  \draw[name path=line, color=gray, semithick] (Rw.south)  -- (Rw.south  |- text.north);
\end{tikzpicture}
\smallskip

The emerging question now is whether an automated translation such as
the one we just sketched preserves treewidth in the following sense:
Given a relational structure $\structure$ of treewidth
$\tw(\structure)$ and an \Lang{mso} sentence $\phi$, can we
mechanically derive a propositional formula~$\psi$ with
$\structure\models\phi$ iff $\psi\in\Lang{sat}$ and
$\tw(\psi)\leq f\big(\tw(\structure)\big)$ for some function
computable $f\colon\mathbb{N}\rightarrow\mathbb{N}$? Consider for
instance the following graph shown on the left (it is ``almost a tree'' and
has treewidth $2$) and the \emph{primal graph} of
$\psi_{\mathrm{3col}}$ obtained using the just sketched
transformation on the right. In this example, the
tree-like structure is preserved, as the treewidth gets
increased by a factor of $3$ and is at most $6$:

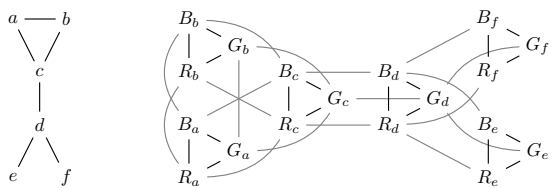
\begin{center}
  \scalebox{0.7}{\begin{tikzpicture}
    \node (1) at (0,   0) {$a$};
    \node (2) at (1,   0) {$b$};
    \node (3) at (.5, -1) {$c$};
    \node (4) at (.5, -2) {$d$};
    \node (5) at (0,  -3) {$e$};
    \node (6) at (1,  -3) {$f$};
    \graph[use existing nodes, edges = {semithick}] {
      1 -- 2 -- 3 -- 1;
      3 -- 4 -- {5, 6};
    };
  \end{tikzpicture}
  \qquad\qquad\qquad
  \begin{tikzpicture}[
    every label/.style = {color=gray},
    rotate = 90,
    yscale = 1.25
    ]
    \newcommand{\triple}[3]{%
      \node (R#3) at ($(#1,#2)+(-.5,0)$)  {$R_#3$};
      \node (B#3) at ($(#1,#2)+(.5,0)$)   {$B_#3$};
      \node (G#3) at ($(#1,#2)+(0,-.75)$) {$G_#3$};
      \draw[semithick] (R#3) -- (G#3);
      \draw[semithick] (R#3) -- (B#3);
      \draw[semithick] (B#3) -- (G#3);
    };
    \triple{0}{0}{a}
    \triple{2}{0}{b}
    \triple{1}{-1.5}{c}
    \triple{1}{-3}{d}
    \triple{0}{-4.5}{e}
    \triple{2}{-4.5}{f}
    \begin{scope}[on background layer]
      \graph[use existing nodes, edges={color=gray, semithick}]{
        Ra --[bend left] Rb;
        Ga -- Gb;
        Ba --[bend left] Bb;
        Ra --[bend right] Rc;
        Ga --[bend right] Gc;
        Ba -- Bc;
        Rc -- Rb;
        Gc --[bend right] Gb;
        Bc --[bend right] Bb;
        Rc -- Rd;
        Gc -- Gd;
        Bc -- Bd;
        Rd -- Re;
        Gd --[bend right] Ge;
        Bd --[bend left] Be;
        Rd --[bend right] Rf;
        Gd --[bend left] Gf;
        Bd -- Bf;
      };
    \end{scope}
  \end{tikzpicture}}
\end{center}

We recap this finding as the following observation: The automated
grounding process from \Lang{mc(mso)} to \Lang{sat} \emph{implies} a
reduction from the 3-coloring problem parameterized by the input's
treewidth to $\Lang{sat}$. We can, thus, derive that the 3-coloring
problem can be solved in time $\tower(1,3k)$ using
Fact~\ref{fact:semiringtw}~--~without actually utilizing Courcelle's
Theorem!

For a second example consider the optimization and counting version of
the dominating set problem: Given a graph $G$, the task is either to
find a minimum-size set $S\subseteq V(G)$ of vertices such that every
vertex is in $S$ or adjacent to vertex in $S$, or to count the number
of such sets. Optimization and counting problems can be
modeled in descriptive complexity by ``moving'' an existential
second-order quantifier (``guessing'' the solution) out of the
sentence and making it a free variable. The task is either to find a
set of minimum size such that the given structure together with this
set is a model of the formula, or to count the number of such sets.
For instance, the following formula
describes that $X$ is a dominating set:
\[
  \phi_{\mathrm{ds}}(X) = \forall x\exists y\suchthat Xx\vee (Exy\wedge Xy).
\]
We will also say that the formula \emph{Fagin-defines} the property
that $X$ is a dominating set. The problem \Lang{\#fd(mso)} asks, given a
relational structure $\structure$ and an \Lang{mso} formula with a
free-set variable~$X$, how many subsets $S$ of the universe of
$\structure$ satisfy $\structure\models\phi(S)$. The
optimization problem \Lang{fd(mso)} gets as additional input an integer $t$
and asks whether there is such a $S$ with $|S|\leq t$. The reduction from \Lang{mc(mso)} to
\Lang{sat} can be extended to a reduction from \Lang{fd(mso)} to
\Lang{maxsat} and from \Lang{\#fd(mso)} to \Lang{pmc}. In order to ground \Lang{fd(mso)}, we
add new indicator variables $X_u$ for the free-variable $X$ and every
element $u$ of $\structure$ (as we did for the second-order
quantifiers). For \Lang{fd(mso)}, we additionally add a soft clause $(\neg X_u)$ for each
of these variables~--~implying that we seek a model that minimizes
$|X|$. We may now again ask: If we mechanically ground $\phi_{\mathrm{ds}}(X)$ on a
structure of bounded treewidth to a propositional formula
$\psi_{\mathrm{ds}}$, what can we say about the treewidth of
$\psi_{\mathrm{ds}}$? 
Unfortunately, not so much. Even if the input has treewidth~$1$, the
primal graph of $\psi_{\mathrm{ds}}$ may become a clique (of
treewidth~$n$):

\begin{center}
  \scalebox{0.7}{\begin{tikzpicture}[baseline=(1.center)]
    \node[circle, inner sep=3pt] (1) at (0,   0) {$a$};
    \node[circle, inner sep=3pt] (2) at (0:1)    {$b$};
    \node[circle, inner sep=3pt] (3) at (45:1)   {$c$};
    \node[circle, inner sep=3pt] (4) at (90:1)   {$d$};
    \node[circle, inner sep=3pt] (5) at (135:1)  {$e$};
    \node[circle, inner sep=3pt] (6) at (180:1)  {$f$};
    \node[circle, inner sep=3pt] (7) at (225:1)  {$g$};
    \node[circle, inner sep=3pt] (8) at (270:1)  {$h$};
    \node[circle, inner sep=3pt] (9) at (315:1)  {$i$};
    \graph[use existing nodes, edges = {semithick}] {
      1 -- {2,3,4,5,6,7,8,9}
    };
  \end{tikzpicture}
  \qquad\qquad
  \begin{tikzpicture}[scale=1.25, baseline=(1.center)]
    \node[circle, inner sep=1pt] (1) at (-2,0)   {$X_a$};
    \node[circle, inner sep=1pt] (2) at (0:1)    {$X_b$};
    \node[circle, inner sep=1pt] (3) at (45:1)   {$X_c$};
    \node[circle, inner sep=1pt] (4) at (90:1)   {$X_d$};
    \node[circle, inner sep=1pt] (5) at (135:1)  {$X_e$};
    \node[circle, inner sep=1pt] (6) at (180:1)  {$X_f$};
    \node[circle, inner sep=1pt] (7) at (225:1)  {$X_g$};
    \node[circle, inner sep=1pt] (8) at (270:1)  {$X_h$};
    \node[circle, inner sep=1pt] (9) at (315:1)  {$X_i$};
    \begin{scope}[on background layer]
      \graph[use existing nodes, edges = {semithick, color=gray}] {
        1 -- {3,5,6,7,9};
        1 --[bend left=15]  2;
        1 --[bend left] 4;
        1 --[bend right] 8;
        2 -- {3,4,5,6,7,8,9};
        3 -- {4,5,6,7,8,9};
        4 -- {5,6,7,8,9};
        5 -- {6,7,8,9};
        6 -- {7,8,9};
        7 -- {8,9};
        8 -- 9;
      };
    \end{scope}
  \end{tikzpicture}}
\end{center}

It follows that we can\emph{not} derive an fpt-algorithm for the
dominating set problem or its counting version by reasoning
about~$\psi_{\mathrm{ds}}$, while we can conclude the fact
from~$\phi_{\mathrm{ds}}$ using appropriate versions of Courcelle's
Theorem.  To summarize, we can naturally describe model-checking,
optimization, and counting problems using monadic second-order
logic. Using Courcelle's Theorem, we can solve all of these problems
in fpt-time on structures of bounded treewidth. Alternatively, we may
ground the \Lang{mso} formulas to propositional logic and solve the
problems using Fact~\ref{fact:semiringtw}. The produced encodings
sometimes preserve the input's structure (as for 3-coloring) and,
thus, themselves serve as proof that the problems lie in
$\Class{FPT}$. However, the input's structure can also get eradicated,
as we observed for the dominating set problem. The present paper is
concerned with the question whether there is a unifying grounding
procedure that maps Fagin-defined \Lang{mso} properties to propositional
logic while preserving the input’s treewidth.

\subparagraph*{Contribution I: Faster Structure-guided Reasoning}
Before we develop a unifying, structure-aware grounding process from
the model-checking problem of monadic second order logic to
propositional logic, we first improve both of the underlying
results. In particular, we remove the logarithmic dependencies on $k$
in top of the tower of Fact~\ref{fact:qsat} and, thus, provide the
first major improvement on \Lang{qbf} upper bounds with respect to treewidth
since 20 years:

\begin{theorem}[QBF Theorem]\label{theorem:qbftw}
  One can solve \Lang{qsat} in time
  $\tower\!\big(\qa(\psi)+1, k + 3.92 \big)$ if a width-$k$ tree decomposition is given.
\end{theorem}

This bound matches the \Lang{eth} lower
bound for \Lang{qsat}:
\begin{fact}[\cite{FichteHP20}]\label{fact:qsat-lb}
  Unless $\Class{ETH}$ fails, \Lang{qsat} cannot be solved in time
  $\tower(\qa(\psi)+1, o(\tw(\psi)))$.
\end{fact}
We will prove Theorem~\ref{theorem:qbftw} fully in the spirit of an
automated reasoning paper by an encoding into \Lang{sat}. In
particular, we will not need any pre-requirements other than
Fact~\ref{fact:semiringtw}. With a similar encoding scheme, we will also
slightly improve on Fact~\ref{fact:pmc}:

\begin{theorem}[PMC Theorem]\label{theorem:pmctw}
  One can solve \Lang{pmc} in time
  $\tower\!\big(2, k + 3.59 \big)$ if a width-$k$ tree decomposition is given.
\end{theorem}
\begin{fact}[\cite{FichteHMTW23}]\label{fact:pmc-lb}
  Unless $\Class{ETH}$ fails, \Lang{pmc} cannot be solved in time
  $\tower(2, o(\tw(\psi)))$.
\end{fact}

\subparagraph*{Contribution II: A SAT Version of Courcelle's Theorem} We
answer the main question of the introduction in the affirmative and
provide a unifying, structure-aware encoding scheme from
properties Fagin-defined with monadic second-order logic to variants
of \Lang{sat}:

\begin{theorem}[A SAT Version of Courcelle's
    Theorem]\label{theorem:sattheorem}
  Assuming that the \Lang{mso} formulas on the left side are in prenex
  normal form and that a width-$k$ tree decomposition is given, there are encodings from \dots
  \begin{enumerate}
    \newcommand\fromtosize[3]{%
    \item\leavevmode\hbox to 0pt{#1\hss}\hspace{1.8cm}to\, \hbox to 0pt{#2\hss}\hspace{1.4cm}of size\, #3%
    }%
  \fromtosize{$\Lang{mc(mso)}$}{$\Lang{sat}$}{$\tower(\qa(\phi),(k+9)|\phi|+3.92)$;}
  \fromtosize{$\Lang{fd(mso)}$}{$\Lang{maxsat}$}{$\tower(\qa(\phi)+1,(k+9)|\phi|+3.92)$;}
  \fromtosize{$\Lang{\#fd(mso)}$}{$\Lang{\#sat}$}{$\tower(\qa(\phi)+1,(k+9)|\phi|+3.92)$.}
  \end{enumerate}
  All encodings of size $\tower(s,t)$ have a treewidth of
  $\mathrm{tower}(s,t)$ and can be computed in linear time with respect to
  their size.
\end{theorem}

In conjunction with Fact~\ref{fact:semiringtw}, the theorem implies
Courcelle's Theorem with sharp bounds on the values on top of the
tower:

\begin{corollary}
  One can solve $\Lang{mc(mso)}$ in time
  $\tower(\qa(\phi)+1,(k+9)|\phi|+3.92)$, and $\Lang{fd(mso)}$ and $\Lang{\#fd(mso)}$
  in time $\tower(\qa(\phi)+2,(k+9)|\phi|+3.92)$ if a width-$k$ tree decomposition is given.
\end{corollary}

Since the reduction~\cite{LiZD04} from \Lang{sat} to integer linear
programming (\Lang{ilp}) is treewidth-preserving and
results in an instance of bounded domain, another consequence of
Theorem~\ref{theorem:sattheorem} is an ``\Lang{ilp} Version of
Courcelle's Theorem'' via the dynamic program for \Lang{ilp}~\cite{JansenK15}.

\subparagraph*{Contribution III: ETH Lower Bounds for the Encoding Size}
Given that we can encode \Lang{mso} definable properties into
\Lang{sat} while preserving the input's treewidth, we may ask next whether we can
improve on the \emph{size} of the encodings. While it is well-known
that incarnations of Courcelle's Theorem have to depend on the input's
treewidth and the formula's size in a non-elementary
way~\cite{AtseriasOliva14a} (and hence, the encodings have to be huge
at some point as well), these insights do not give us precise bounds
on achievable encoding sizes. 

\begin{theorem}[ETH Lower Bound]\label{theorem:encoding-lb}
  Under $\Class{ETH}$, there is no \Lang{sat} encoding for
  $\Lang{mc(mso)}$ of size
  $\tower(\qa(\phi)-2,o(\tw(\structure)))$  that can be computed in
  this time.
\end{theorem}

We can make the lower bound a bit more precise in the following sense:
The value at the top of the tower actually does not just depend on the
treewidth $\tw(\structure)$, but on the product of the treewidth and
the \emph{block size} $\bs(\phi)$ of the sentence $\phi$. The block size of a
formula is the maximum number of consecutive quantifiers of the same
type. 
\begin{theorem}[Trade-off Theorem]\label{theorem:tradeoff}
  Under $\Class{ETH}$,
  there is no $\Lang{sat}$ encoding for
  $\Lang{mc(mso)}$ of size $\tower(\qa(\phi)-2,o(\tw(\structure)\bs(\phi)))$ that can be
  computed within this time.
\end{theorem}

\subsection{Related Work}
The concept of treewidth was discovered multiple times. The name was
coined in the work by Robertson and Seymour~\cite{RobertsonS86}, while
the concept was studied by Arnborg and Proskurowski~\cite{ArnborgP85}
under the name partial $k$-trees simultaneously. However, treewidth
was discovered even earlier by Bertelè and
Brioschi~\cite{bertele1972nonserial}, and independently by
Halin~\cite{halin1976s}.  Courcelle's Theorem was proven in a series
of articles by Bruno Courcelle~\cite{Courcelle90}, see also the
textbook by Courcelle and Engelfriet for a detailed
introduction~\cite{CourcelleE12}. The expressive power of monadic
second-order logic was studied before, prominently by Büchi who showed
that \Lang{mso} over strings characterizes the regular
languages~\cite{Buechi1960}. Related to our treewidth-aware reduction
from \Lang{mc(mso)} to \Lang{sat} is the work by Gottlob, Pichler, and
Wei, who solve \Lang{mc(mso)} using monadic
Datalog~\cite{GottlobPW10}; and the work of Bliem, Pichler, and
Woltran, who solve it using \Lang{asp}~\cite{BliemPW13}.

\subsection{Structure of this Article}
We provide preliminaries in the next section, prove Theorem~\ref{theorem:qbftw}
and~\ref{theorem:pmctw} in Section~\ref{section:qbf-pmc}, and
establish a \Lang{sat} version of Courcelle's Theorem in
Section~\ref{section:mso-to-sat}. The technical details of the latter can
be found in Appendix~\ref{section:mso-to-sat-details}. We extend the result to
Fagin-definable properties in
Section~\ref{section:mso-optimization-and-counting} and provide
corresponding $\Class{ETH}$ lower bounds in
Section~\ref{section:lowerbounds}. We conclude and provide pointers
for further research in the last section, which also contains an
overview table of this article's results. 

\tcsautomoveaddto{main}{
  \clearpage
  \appendix
  \section{Technical Appendix}
  In the following, we provide the proofs omitted in the main text. In
  each case, the claim of the theorem or lemma is stated once more for
  the reader's convenience. 
}

%
%
\section{Preliminaries: Background in Logic and Structural Graph Theory}
\label{section:preliminaries}

We use the notation of Knuth~\cite{Knuth16} and consider 
\emph{propositional formulas} in conjunctive normal form (\Lang{cnf}s)
like
\(
\psi =
(x_1\vee \neg x_2\vee \neg x_3)
\wedge (\neg x_1\vee x_4\vee\neg x_5)
\wedge (x_2)
\wedge (x_6)
\)
as \emph{set of sets}
\(
\{
\{x_1,\neg x_2, \neg x_3\},
\{\neg x_1, x_4, \neg x_5\},
\{x_2\},
\{x_6\}
\}
\).  We denote the sets of variables, literals, and clauses of
$\psi$ as $\vars(\psi)$, $\lits(\psi)$, and $\clauses(\psi)$. 
A \emph{(partial) assignment} is a
subset $\beta\subseteq \lits(\psi)$ such that $|\{x,\neg x\}\cap
\beta|\leq 1$ for all $x\in\vars(\psi)$, that is, a set of literals that does not contain both
polarities of any variable. We use $\beta\assignment\vars(\psi)$ to denote partial assignments.
The formula \emph{conditioned under a
  partial assignment} $\beta$ is denoted by $\psi|\beta$ and obtained
by removing all clauses from $\psi$ that
contain a literal $l\in\beta$ and by removing all literals $l'$ with
$\neg l'\in\beta$ from the remaining clauses. 
A assignment is \emph{satisfying} for a \Lang{cnf} $\psi$ if
$\psi|\beta=\emptyset$, and it is \emph{contradicting} if
$\emptyset\in\psi|\beta$. A \Lang{dnf} is a disjunction of
conjunctions, i.e., a set of terms. We use the same notations as for
\Lang{cnf}s, however, in $\psi|\beta$ we delete
terms that contain a literal that appears negated in $\beta$ and remove the literals in
$\beta$ from the remaining terms. Hence, $\beta$ is \emph{satisfying} if $\emptyset\in\psi|\beta$, and \emph{contradicting} if $\psi|\beta=\emptyset$.

The \emph{model counting problem} asks to
compute the number of satisfying assignments of a \Lang{cnf} and
is denoted by $\Lang{\#sat}$. In \emph{projected model
  counting} (\Lang{pmc}) we count the number of models that are
not identical on a given set of variables.
In the \emph{maximum satisfiability problem} (\Lang{maxsat})
we partition the clauses of $\psi$ into a set $\hard(\psi)$ of \emph{hard clauses} and
a set $\soft(\psi)$ of weighted \emph{soft clauses}, i.e., every clause $C\in\soft(\psi)$ comes
with a \emph{weight} $w(C)\in\mathbb{Q}$. The formula is then called a \Lang{wcnf} and
the goal is to find under all assignments $\beta\assignment\vars(\psi)$
with $\hard(\psi)|\beta=\emptyset$ the one that maximizes
$\sum_{C\in\soft(\psi), \{C\}|\beta=\emptyset}w(c)$.
In a fully \emph{quantified Boolean formula} (a \Lang{qbf}, also called a \emph{second-order
propositional sentence}) all variables are bounded by existential or
universal quantifiers. Throughout the paper we assume that \Lang{qbf}s
are in prenex normal form, meaning that all quantifers appear in the
front of a quantifier-free formula called the \emph{matrix}. As is
customary, we assume that the matrix is a \Lang{cnf} if the last
(i.e., most inner) quantifier is existential, and
a \Lang{dnf} otherwise. A \Lang{qbf} is \emph{valid} if it
evaluates to true (see Chapter~29--31
in~\cite{HandbookSAT}). Define \Lang{qsat} to be the problem of deciding whether
a given \Lang{qbf} is valid.

\subsection{Descriptive Complexity}  A
\emph{vocabulary} is a finite set
$\tau=\{R_1^{a_1},R_2^{a_2},\dots,R_\ell^{a_\ell}\}$ of relational
symbols $R_i$ of arity $a_i$. A (finite, relational)
\emph{$\tau$-structure} $\structure$ is a tuple
$\big(U(\structure), R_1^{\structure}, R_2^{\structure},\dots,
R_\ell^{\structure}\big)$ with \emph{universe} $U(\structure)$ and
\emph{interpretations} $R^{\structure}_i\subseteq U(\structure)^{a_i}$. The \emph{size}
of $\structure$ is
$|\structure|=|U(\structure)|+\sum_{i=1}^\ell a_i\cdot|R_i^\structure|$.
We denote the
\emph{set of all $\tau$-structures} by $\struc[\tau]$~--~e.g.,
$\struc[\{E^2\}]$ is the set of directed graphs.

Let $\tau$ be a vocabulary and $x_0,x_1,x_2,\dots$ be an infinite
repertoire of first-order variables. The \emph{first-order language}
$\lang(\tau)$ is inductively defined, where the \emph{atomic formulas}
are the strings $x_i=x_j$ and $R_i(x_{1},\dots,x_{a_i})$ for
relational symbols $R_i\in\tau$. If $\alpha,\beta\in\lang(\tau)$ then
so are $\neg(\alpha)$, $(\alpha\wedge\beta)$, and
$\exists x_i(\alpha)$. A variable that appears next to $\exists$ is
called \emph{quantified} and \emph{free} otherwise. We denote a
formula $\phi\in\lang(\tau)$ with $\phi(x_{i_1},\dots,x_{i_q})$ if
$x_{i_1},\dots,x_{i_q}$ are precisely the free variables in $\phi$. A
formula without free variables is called a \emph{sentence.} As
customary, we extend the language of first-order logic by the usual
abbreviations, e.g.,
$\alpha\rightarrow\beta\equiv \neg\alpha\vee\beta$ and
$\forall x_i(\alpha)\equiv\neg\exists x_i(\neg\alpha)$. To increase
readability, we will use other lowercase Latin letters for variables
and drop unnecessary braces by using the usual operator precedence
instead. Furthermore, we use the \emph{dot notation} in which we place
a ``$\suchthat$'' instead of an opening brace and silently close it at
the latest syntactically correct position. A $\tau$-structure
$\structure$ is a \emph{model} of a sentence $\phi\in\lang(\tau)$,
denoted by $\structure\models\phi$, if it evaluates to true under the
semantics of quantified propositional logic while interpreting equality
and relational symbols as specified by the structure. For instance,
$\phi_{\mathrm{undir}}=\forall x\forall y\suchthat Exy\rightarrow Eyx$
over $\tau=\{E^2\}$ describes the set of undirected graphs, and we
have $\picocycleundir\models\phi_{\mathrm{undir}}$ and
$\picocycle\not\models\phi_{\mathrm{undir}}$.

We obtain the language of \emph{second-order logic} by allowing
quantification over relational variables of arbitrary arity, which we
will denote by uppercase Latin letters. A relational variable is said
to be \emph{monadic} if its arity is one. A \emph{monadic
  second-order} formula is one in which all
quantified relational variables are monadic. The set of all
such formulas is denoted by \Lang{mso}. The \emph{model checking
  problem} for a vocabulary $\tau$ is the set
$\Lang{mc}_\tau(\Lang{mso})$ that contains all pairs $(\structure,\phi)$ of
$\tau$-structures $\structure$ and \Lang{mso} sentences $\phi$ with
$\structure\models\phi$. Whenever $\tau$ is not relevant (meaning that a result
holds for all fixed $\tau$), we will refer to the problem as
$\Lang{mc}(\Lang{mso})$. We note that in the literature there is often a
distinction between $\Lang{mso}_1$- and $\Lang{mso}_2$-logic, which
describes the way the input is encoded~\cite{HlinenyOSG08}. Since we allow arbitrary
relations, we do not have to make this distinction.

\subsection{Treewidth and Tree Decompositions}  While we consider
graphs $G$ as relational structures~$\graphstructure$ as discussed in the
previous section, we also use common graph-theoretic terminology and
denote with $V(G)=U(\graphstructure)$ and $E(G)=E^{\graphstructure}$ the
vertex and edge set of $G$. Unless stated otherwise, graphs in this
paper are \emph{undirected} and we use the natural set
notations and write, for instance,
$\{v,w\}\in E(G)$. The
\emph{degree} of a vertex is the number of its neighbors. 
A \emph{tree decomposition} of $G$ is a pair $(T,\bag)$ in
which $T$ is a tree (a connected graph without cycles) and
$\bag\colon V(T)\rightarrow 2^{V(G)}$ a function with the following
two properties:

\begin{enumerate}
\item for every $v\in V(G)$ the set $\{\, x\mid v\in\bag(x)\,\}$ is
  non-empty and connected in $T$;
\item for every $\{u,v\}\in E(G)$ there is at least one node $x\in V(T)$ with
  $\{u,v\}\subseteq \bag(x)$.
\end{enumerate}

The \emph{width} of a tree decomposition is the maximum size of its
bags minus one, i.e., $\width(T,\bag)=\max_{x\in
  V(T)}|\bag(x)|-1$. The \emph{treewidth} $\tw(G)$ of a graph $G$ is
the minimum width any tree decomposition of $G$ must have. We do not
require additional properties of tree decompositions, but we assume
that $T$ is rooted at a $\rootOf(T)\in V(T)$ and, thus, that
nodes $t\in V(T)$ may have a $\parent(t)\in V(T)$ and
$\children(t)\subseteq V(T)$. Without loss of generality, we may also assume
$|\children(t)|\leq 2$.

\begin{example}\label{example:tw}
  The treewidth of the Big Dipper constellation (as graph shown on the
  left) is at most two, as proven by the tree decomposition on the right:

  \null\hfill\scalebox{0.7}
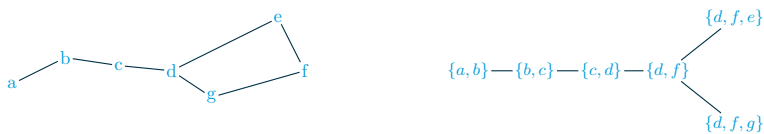
{\begin{tikzpicture}[baseline=(f),
      star/.style = {
        color     = fg,
        inner sep = 1pt
      }
    ]
    \node[star] (a) at (0,-.25)   {a};
    \node[star] (b) at (1,.25)    {b};
    \node[star] (c) at (2,.1)     {c};
    \node[star] (d) at (3,0)      {d};
    \node[star] (e) at (5,1)      {e};
    \node[star] (f) at (5.5,0)    {f};
    \node[star] (g) at (3.75,-.5) {g};

    \graph[use existing nodes, edges = {semithick}]{
      a -- b -- c -- d -- e -- f -- g -- d;
    };
  \end{tikzpicture}}
  \hfill
  \scalebox{0.7}{\begin{tikzpicture}[rotate=-90, yscale=1.25,
        baseline=(b1),
      bag/.style = {
        color     = fg,
        inner sep = 1pt,
        font      = \small
      }
    ]
    \node[font=\small, bag] (b1) at (0,0)  {$\{a,b\}$};
    \node[font=\small, bag] (b2) at (0,1)  {$\{b,c\}$};
    \node[font=\small, bag] (b3) at (0,2)  {$\{c,d\}$};
    \node[font=\small, bag] (b4) at (0,3)  {$\{d,f\}$};
    \node[font=\small, bag] (b5) at (-1,4) {$\{d,f,e\}$};
    \node[font=\small, bag] (b6) at (1,4)  {$\{d,f,g\}$};
    
    \graph[use existing nodes, edges = {semithick}]{
      b1 -- b2 -- b3 -- b4 -- {b5, b6}
    };
  \end{tikzpicture}}\hfill\null
\end{example}

\subsection{Treewidth of Propositional Formulas and Relational
  Structures} The definition of treewidth can be lifted to other
objects by associating a graph to them. The most common graph for
\Lang{cnf}s (or \Lang{dnf}s) $\psi$ is the \emph{primal graph}
$G_\psi$, which is the graph on vertex set $V(G_\psi)=\vars(\psi)$
that connects two vertices by an edge if the corresponding variables
appear together in a clause. We then define
$\tw(\psi)\coloneq\tw(G_\psi)$ and refer to a tree decomposition of
$G_\psi$ as one of $\psi$. Note that other graphical representations
lead to other definitions of the treewidth of propositional
formulas. A comprehensive listing can be found in the \emph{Handbook
of Satisfiability}~\cite[Chapter~17]{HandbookSAT}. A \emph{labeled
tree decomposition} $(T,\bag,\baglabel)$ extends a tree decomposition
with a mapping $\baglabel\colon V(T)\rightarrow 2^{\psi}$ (i.e., a
mapping from the nodes of~$T$ to a subset of the clauses (or terms) of
$\psi$) such that for every clause (or term) $C$ there is exactly one
$t\in V(T)$ with $C\in\baglabel(t)$ that contains all variables
appearing in $C$. It is easy to transform a tree decomposition
$(T,\bag)$ into a labeled one $(T,\bag,\baglabel)$ by traversing the
tree once's and by duplicating some bags. Hence, we will assume
throughout this article that all tree decompositions are labeled.

A similar approach can be used to define tree decompositions of
arbitrary structures: The \emph{primal graph} $G_\structure$
of a structure $\structure$, in this context also called the
\emph{Gaifman graph}, has as vertex set the universe of $\structure$,
i.e., $V(G_\structure)=U(\structure)$, and
contains an edge $\{u,v\}\in E(G_\structure)$ iff
$u$ and $v$ appear together in some tuple of $\structure$. As before,
we define
$\tw(\structure)\coloneq\tw(G_\structure)$. One can alternatively define the
concept of tree decompositions directly over relational structures,
which leads to the same definition~\cite{FlumG06}.

%
%
\section{New Upper Bounds for Second-Order Propositional Logic}\label{section:qbf-pmc}
\tcsautomoveaddto{main}{\subsection{Proofs for Section~\ref{section:qbf-pmc}}}

Central to our reductions are
treewidth-preserving encodings from \Lang{qsat} to \Lang{sat} and
from \Lang{pmc} to \Lang{\#sat}. These
encoding establishes new proofs of Chen's Theorem~\cite{Chen04} and
the theorem by Fichte et al.~\cite{FichteHMTW23}, and
improve the dependencies on $k$ in the tower of
Fact~\ref{fact:qsat} and~\ref{fact:pmc}.

\subsection{Treewidth-Aware Encodings from QSAT to SAT}

We use a quantifier elimination
scheme that eliminates the most-inner quantifier block at the cost of
introducing $O(2^k|\phi|)$ new variables while increasing
the treewidth by a factor of $12\cdot 2^k$. Let first $\phi=Q_1
S_1\dots \forall_\ell S_\ell\suchthat \psi$ be the given \Lang{qbf},
in which $\psi$ is a \Lang{dnf}.  Let further $(T,\bag,\baglabel)$ be
the given labeled width-$k$ tree decomposition of $\phi$. We describe
an encoding into a \Lang{qbf}, in which the last quantifier block
$Q_\ell S_\ell$ gets replaced by new variables in $S_{\ell-1}$.

We have to encode the fact that for an assignment on
$\bigcup_{i=1}^{\ell-1}S_i$ \emph{all} assignments to~$S_\ell$
satisfy~$\psi$, i.\,e., at least \emph{one} term in $\psi.$ For that
end, we introduce auxiliary variables for every term
$d\in\termsx(\psi)$ and any partial assignment $\alpha$ of the
variables in $S_\ell$ that also appear in the bag that contains
$d$. More precisely, let $\baglabel^{-1}(d)$ be the node in $V(T)$
with $d\in\baglabel(t)$ and let
$\alpha\assignment\bag(\baglabel^{-1}(d))\cap S_\ell$ be an assignment
of the variables of the bag that are quantified by~$Q_\ell$. We
introduce the variable $\sat^{\alpha}_d$ that indicates that this
assignment satisfies~$d$:
\begin{align*}
  &\bigwedge_{d\in\termsx(\psi)}
  \bigwedge_{\substack{\alpha\assignment\bag(\baglabel^{-1}(d))\cap S_\ell\\\{d\}\mid\alpha\neq\emptyset}}
  \Big[\,
    \sat_d^{\alpha}\leftrightarrow
    \bigwedge_{\mathclap{x\in \lits(\{d\}\mid\alpha)}}x
    \,\Big],\tag*{(1)}
  &&\comment{// $\alpha$ may satisfy $d$}\\        
  &\bigwedge_{d\in\termsx(\psi)}
  \bigwedge_{\substack{\alpha\assignment\bag(\baglabel^{-1}(d))\cap S_\ell\\\{d\}|\alpha=\emptyset}}
  \Big[\,
    \neg \sat_d^{\alpha}
    \,\Big].
  &&\comment{// $\alpha$ falsifies $d$}
  \tag*{(2)}
\end{align*}
We have to track whether $\psi$ can be satisfied by a local assignment
$\alpha$. For every $t\in V(T)$ and every
$\alpha\assignment\bag(t)\cap S_\ell$ we introduce a variable
$\sat^{\alpha}_{\leq t}$ that indicates that $\alpha$ can be extended
to a satisfying assignment for the subtree rooted at $t$. Furthermore,
we create variables $\sat^{\alpha}_{<t,t'}$ for $t'\in\children(t)$
that propagate the information about satisfiability along the tree
decomposition. That is, $\sat^{\alpha}_{<t,t'}$ is set to true if
there is an assignment $\beta\assignment\bag(t')\cap S_\ell$ that can
be extended to a satisfying assignment and that is compatible with
$\alpha$:
\begin{align*}
  &\comment{// Either there is a term satisfing the bag or we can propagate:}\\
  &\bigwedge_{t\in V(T)}
  \bigwedge_{\alpha\assignment\bag(t)\cap S_\ell}
  \Big[\,
    \sat^{\alpha}_{\leq t}\leftrightarrow\bigvee_{\mathclap{d\in\baglabel(t)}}\sat_d^{\alpha}\vee\bigvee_{\mathclap{t'\in\children(t)}}\sat^{\alpha}_{<t,t'}
    \,\Big],\tag*{(3)}\\[2ex]
  &\comment{// Propagate satisfiability:}\\
  &\bigwedge_{t\in V(T)}
  \bigwedge_{\alpha\assignment\bag(t)\cap S_\ell}
  \bigwedge_{t'\in\children(t)}      
  \Big[\,
    \sat^{\alpha}_{< t,t'}\leftrightarrow
    \bigwedge_{\mathclap{\substack{\beta\assignment\bag(t')\cap S_\ell\\\beta\cap\lits(\bag(t))=\alpha\cap\lits(\bag(t'))}}}\sat^{\beta}_{\leq t'}
    \,\Big].\tag*{(4)}
\end{align*}
Finally, since $Q_\ell=\forall$, we need to ensure that for all
possible assignments of $S_\ell$ there is at least one term that gets
satisfied. Since satisfiability gets propagated to the root of the tree
decomposition by the aforementioned constraint, we can enforce this
property with:
\[
\bigwedge_{\alpha\assignment\bag(\rootOf(T))\cap S_\ell}\hspace{-1em}\sat^{\alpha}_{\leq\rootOf(T)}.\tag*{(5)}
\]
  
The following lemma observes the correctness of the construction, and
the subsequent lemma handles the case $Q_\ell=\exists$.

\begin{lemma~}\label{lemma:qbf-eliminate-forall}
  There is an algorithm that, given a \Lang{qbf} $\phi=Q_1 S_1\dots
  \exists_{\ell-1} S_{\ell-1}\forall_\ell S_\ell\suchthat \psi$ and a
  width-$k$ tree decomposition of $G_\phi$, outputs in time $O^*(2^k)$
  a \Lang{qbf} $\phi'=Q_1 S_1\dots \exists_{\ell-1}
  S'_{\ell-1}\suchthat \psi'$ and a width-$(12\cdot2^k)$ tree
  decomposition of $G_{\phi'}$ such that $\phi$ is valid iff $\phi'$
  is valid.
\end{lemma~}

\begin{proof~}
  To prove the correctness of the encoding first observe that the
  encoding can clearly be written as \Lang{cnf}. Now take any
  assignment~$\beta\sqsubseteq S_1\cup \ldots \cup S_{\ell-1}$ such
  that for all~$\gamma\sqsubseteq S_\ell$ we have that
  $\phi|(\beta\cup \gamma)$ evaluates to true. From this we construct
  $\gamma'\sqsubseteq (\var(\phi')\setminus \var(\phi))$ and show that
  $\phi'|(\beta\cup \gamma')$ evaluates to true.  More precisely, we
  let
  \begin{align*}
    \gamma_{0} =\,
    &\{\,
    \sat_d^\alpha, \sat_{\leq t}^\alpha \mid \alpha \sqsubseteq {\bag(t)},  d\in\baglabel(t), \{d\}|(\alpha \cup \beta)\neq\emptyset
    \,\}\\
    \cup\,
    &\{\,
    \neg \sat_d^\alpha \mid \alpha \sqsubseteq  \bag(t), \{d\}|(\alpha \cup  \beta)=\emptyset
    \,\}.
  \end{align*}
  For every
  node~$t\in V(T)$ and assignment~$\alpha\sqsubseteq\bag(t)$, we
  define the set of compatible assignments for a
  node~$t'\in\children(t)$ as
  \begin{align*}
    \compat(\alpha, t, t')&=\{\,\beta\sqsubseteq \bag(t'), \alpha\cap\lits(\bag(t'))\\&=\phantom{\{\,}\beta\cap\lits(\bag(t))\,\}.
  \end{align*}
  Inductively
  we define for every~$1\leq i\leq |V(T)|$ the set
  \begin{align*}
    \gamma_i &=
    \gamma_{i-1}
    \cup
    \{\,
    \sat_{\leq t}^\alpha, \sat_{<t,t'}^\alpha
    \mid t\text{ in }T,\\
    &\qquad  \{\,\sat_{\leq t'}^\beta \mid \beta\in\compat(\alpha, t,t')\,\}\subseteq \gamma_{i-1}
    \,\}.
  \end{align*}
  We
  define~$\gamma'=\gamma_{|V(T)|} \cup \{\,\neg x \mid x\in
  \var(\phi')\setminus \gamma_{|V(T)|}\,\}$, which assigns remaining
  variables to zero.  It remains to show that we have
  $\phi'|(\beta\cup \gamma')\neq\emptyset$.  Suppose towards a
  contradiction that $\phi'|(\beta\cup \gamma')$ does not evaluate to
  false.  The reason for that can only be due to Equation~(5), as
  obviously Equations~(1)--(4) are satisfied by the construction
  of~$\gamma'$.  However, then, by construction of~$\gamma'$, one can
  follow down assignments~$\alpha\sqsubseteq \bag(\rootOf(T))$
  with~$\neg \sat_{\leq \rootOf(T)}^\alpha \in\gamma'$, i.\,e., for every
  node~$t'\in\children(t)$, we collect a compatible
  assignment~$\alpha_{t'}\in\compat(\alpha, t, t')$,
  where~$\neg \sat_{\leq t'}^\alpha \in\gamma'$. By construction, such an
  assignment~$\alpha_{t'}$ has to exist.

  Following these
  assignments to the leaf nodes of~$T$, we can combine the thereby
  obtained assignments, which leads to an assignment~$\alpha'$ such
  that $\phi|(\beta\cup \alpha')\neq\emptyset$.  Consequently,
  $\alpha'$ is a witness that contradicts our assumption that
  $\phi|(\beta\cup \gamma)=\emptyset$ for any~$\gamma\sqsubseteq
  S_\ell$.

  For the other direction take any $\alpha\sqsubseteq \var(\phi')$
  with $\phi'|\alpha$ being valid, i.\,e., $\phi'|\alpha=\emptyset$.
  We show that for $\alpha'=\alpha\cap \lits(S_1\cup \ldots \cup
  S_{\ell-1})$, $\phi|\alpha'$ is valid as well, i.\,e.,
  $\phi|\alpha'\neq \emptyset$.  Suppose towards a contradiction that
  $\phi|\alpha'$ is invalid. Then, there exists an assignment
  $\beta'\sqsubseteq S_\ell$ such that $\phi|(\alpha' \cup \beta')$
  evaluates to false.  Since~$\beta'$ witnesses the invalidity of
  $\phi|\alpha'$, any assignment $\beta\sqsubseteq\var(\phi')$
  containing $\alpha$ (i.\,e., $\alpha\subseteq \beta$) either
  invalidates Equation~(5) or one of~(1)--(4). This contradicts that
  $\phi'|\alpha$ is valid and we demonstrated a bijective relationship
  between the satisfying assignments of~$\psi$ over variables $S_1\cup
  \ldots \cup S_{\ell-1}$ and the satisfying assignments of~$\psi'$
  over $S_1\cup \ldots \cup S_{\ell-1}$. We conclude that $\phi$ is
  valid if and only if $\phi'$ is valid.

  For the bound on the treewidth we observe that the last quantifier block
  now contains more variables (i.\,e., it quantifies the auxiliary
  variables), but the sets $S_i$ for $i<\ell-1$ are unchanged. Clearly, $\phi'$ contains
  $O(2^k|\phi|)$ new variables. To see that the treewidth is increased
  by at most $12\cdot 2^{k}$, observe that there is a tree decomposition in
  which every bag contains at most one term, i.\,e., we only add the
  variables $\sat_d^{\alpha}$ (which are $2^{k+1}$) and the propagation
  variables $\sat^{\alpha}_{\leq t}$ (which are $2^{k+1}$) and
  $\sat^{\alpha}_{<t,t'}$ (which are at most $2\cdot 2^{k+1}$,
  because~$t$ has at most two children), as well as
  $\sat^{\beta}_{\leq t'}$ (which are again $2\cdot 2^{k+1}$) to the bag. Hence, the
  treewidth grows by at most $2^{k+1}+2^{k+1}+2\cdot 2^{k+1}+2\cdot 2^{k+1}=6\cdot
  2^{k+1}=12\cdot 2^k$. It straightforward to obtain a tree
  decomposition with this bound from the given one.
\end{proof~}

\begin{lemma~}\label{lemma:qbf-eliminate-exists}
  There is an algorithm that, given a \Lang{qbf} $\phi=Q_1 S_1\dots
  \forall_{\ell-1} S_{\ell-1}\exists_\ell S_\ell\suchthat \psi$ and a
  width-$k$ tree decomposition of $G_\phi$, outputs in time $O^*(2^k)$
  a \Lang{qbf} $\phi'=Q_1 S_1\dots \forall_{\ell-1}
  S'_{\ell-1}\suchthat \psi'$ and a width-$(12\cdot2^k)$ tree
  decomposition of $G_{\phi'}$ such that $\phi$ is valid iff $\phi'$
  is valid.
\end{lemma~}

\begin{proof~}
  We need to encode that for an assignment on
  $\bigcup_{i=1}^{\ell-1}S_i$ \emph{one} assignment to~$S_\ell$
  satisfy $\psi$, i.\,e., \emph{all} clauses in~$\psi.$ Again, we
  introduce auxiliary variables $\sat_c^{\alpha}$ for every clause
  $c\in\clauses(\psi)$ and every partial assignment
  $\alpha\assignment\bag(\baglabel^{-1}(c))\cap S_\ell$ that indicates
  that $c$ is satisfied in its bag:
  \begin{align*}
    &\bigvee_{c\in\clauses(\psi)}
    \bigvee_{\substack{\alpha\assignment\bag(\baglabel^{-1}(c))\cap S_\ell\\\{c\}\mid\alpha\neq\emptyset}}
    \neg\Big[\,
     \sat_c^{\alpha}\leftrightarrow
    \bigvee_{\mathclap{x\in \lits(\{c\}\mid\alpha)}}x
    \,\Big],\tag*{(6)}\\
   &\bigvee_{c\in\clauses(\psi)}
     \bigvee_{\substack{\alpha\assignment\bag(\baglabel^{-1}(c))\cap S_\ell\\\{c\}\mid\alpha=\emptyset}}
   \neg\Big[\,
   \sat_c^{\alpha}
   \,\Big].\tag*{(7)}
  \end{align*}
  We need variables $\sat^{\alpha}_{\leq t}$ and
  $\sat^{\alpha}_{< t, t'}$ to propagate information along the tree
  decomposition. The corresponding constraints
  become:  
  \begin{align*}
    &\comment{// Either there is a clause satisfing the bag or we can propagate:}\\
    &\bigvee_{t\in V(T)}
    \bigvee_{\alpha\assignment\bag(t)\cap S_\ell}
    \neg\Big[\,
    \sat^{\alpha}_{\leq t}\leftrightarrow\bigwedge_{\mathclap{c\in\baglabel(t)}}\sat_c^{\alpha}\wedge\bigwedge_{\mathclap{t'\in\children(t)}}\sat^{\alpha}_{<t,t'}
    \,\Big],\tag*{(8)}\\
    &\comment{// Propagate satisfiability:}\\
    &\bigvee_{t\in V(T)}
      \bigvee_{\alpha\assignment\bag(t)\cap S_\ell}
      \bigvee_{t'\in\children(t)}      
      \neg\Big[\,
      \sat^{\alpha}_{< t,t'}\leftrightarrow
      \bigvee_{\mathclap{\substack{\beta\assignment\bag(t')\cap S_\ell\\\beta\cap\lits(\bag(t))=\alpha\cap\lits\bag(t')}}}\sat^{\beta}_{\leq t'}
      \,\Big].\tag*{(9)}
  \end{align*}
  
  Since $Q_\ell=\exists$, we need to ensure that there
   is an assignments of~$S_\ell$ such that all clauses are satisfied. Satisfiability gets propagate to the root of
   the tree decomposition as in the previous case:
  \[
    \bigvee_{\alpha\assignment\bag(\rootOf(T))\cap S_\ell}\hspace{-1em}\sat^{\alpha}_{\leq\rootOf(T)}.\tag*{(10)}
  \]
  Note that in the constraints (6)--(9) the semantic changed, because
  we now evaluate a \Lang{cnf} (this happens within the brackets). All
  these constraints are negated because the auxiliary variables are
  quantified universally, i.\,e., if one of the propagation
  constraints is falsified we can ``skip'' it, because the
  corresponding assignment is trivially true. In contrast,
  constraint~(10), which states that one of the assignments of the
  root must be satisfied, is not negated, as this corresponds to the
  actual satisfaction of the previous \Lang{cnf}. Correctness and the
  bound on the treewidth of this encoding follows analogously to Lemma~\ref{lemma:qbf-eliminate-forall}.
\end{proof~}

\begin{proof}[Proof of Theorem~\ref{theorem:qbftw}]
  The theorem follows by exhaustively applying
  Lemma~\ref{lemma:qbf-eliminate-forall} and
  Lemma~\ref{lemma:qbf-eliminate-exists} until a \Lang{cnf} is
  reached. The price for removing one alternation are $O(2^k|\phi|)$
  new variables and an increase of the treewidth by a factor of $12\cdot
  2^{k}$. Hence, after removing one quantifier block we have a
  treewidth of $12\cdot 2^{k}\leq 2^{k+\log 12}$, after two we have
  $12\cdot 2^{2^{k+\log 12}}\leq 2^{2^{k+\log 12}+\log 12}$, after
  three we then have $2^{2^{2^{k+\log 12}+\log 12}+\log 12}$; and so
  on. We can bound all the intermediate ``$+\log 12$'' by adding a
  ``$+1$'' on top of the tower, leading to a bound on the treewidth of
  \(
  \mathrm{tower}(\qa(\phi),k+\log 12+1)\leq
  \mathrm{tower}(\qa(\phi),k+4.59).
  \)
  In fact, we can bound the
  top of the tower even tighter by observing $\log 12\leq 3.59$ and
  guessing $3.92$ as a fix point. Inserting yields
  $3.59+2^{3.59+k}\leq 2^{3.92+k}$ and $2^{3.92+2^{3.59+k}}\leq2^{2^{3.92+k}}$.
  Consequently, we can
  bound the treeewidth of the encoding by
  $\toweronly(\qa(\phi),k+3.92)$ and the size by
  $\tower(\qa(\phi),k+3.92)$. 
\end{proof}

%
%
\subsection{Treewidth-Aware Encodings from PMC to \#SAT}

Recall that the input for $\Lang{pmc}$ is a \Lang{cnf}
$\psi$ and a set $X\subseteq\vars(\psi)$. The task is to count the
assignments $\alpha\assignment X$ that can be extended to models
$\alpha^*\assignment\vars(\psi)$ of $\psi$. We can also think of a
formula $\psi(X) = \exists Y\suchthat\psi'(X,Y)$ with free variables
$X$ and existential quantified variables~$Y$ ($\psi'$ is
quantifier-free), for which we want to count the assignments to $X$
that make the formula satisfiable. The idea is to rewrite \( \psi(X) =
\exists Y\suchthat\psi'(X,Y) \equiv \exists X\exists
Y\suchthat\psi'(X,Y), \) and to use a similar encoding as in the proof
of Lemma~\ref{lemma:qbf-eliminate-exists} to remove the second quantifier.

In detail, we add a variable $\sat^{\alpha}_c$ for every clause
$c\in\clauses(\psi)$ and every assignment of the corresponding bag
$\alpha\assignment\bag(\baglabel^{-1}(c))\cap Y$. The semantic of this
variable is that the clause $c$ is satisfiable under the partial
assignment $\alpha$. We further add the propagation variables
$\sat^{\alpha}_{\leq t}$ and $\sat^{\alpha}_{< t, t'}$ for all $t\in
V(T)$, $t'\in\children(t)$, and
$\alpha\assignment\bag(\baglabel^{-1}(c))\cap Y$. The former indicates
that the assignment $\alpha$ can be extended to a satisfying
assignment of the subtree rooted at $t$; the later propagates partial
solutions from children to parents within the tree decomposition:
\begin{align*}
  &\comment{// $\alpha$ may satisfy $c$:}\\
  &\bigwedge_{c\in\clauses(\psi)}
  \bigwedge_{\substack{\alpha\assignment\bag(\baglabel^{-1}(c))\cap Y\\\{c\}\mid\alpha\neq\emptyset}}
  \Big[\,
    \sat_c^{\alpha}\leftrightarrow
    \bigvee_{\mathclap{\ell\in\lits(\{c\}\mid\alpha)}}\ell
    \,\Big],\tag*{(1)}
\end{align*}
\begin{align*}
  &\comment{// $\alpha$ satisfies $c$:}\\
  &\bigwedge_{c\in\clauses(\psi)}
  \bigwedge_{\substack{\alpha\assignment\bag(\baglabel^{-1}(c))\cap Y\\\{c\}\mid\alpha=\emptyset}}
  \Big[\,
    \sat_c^{\alpha}
    \,\Big].\tag*{(2)}
\end{align*}
\begin{align*}
  &\comment{// Either there is a clause satisfying the bag or we can propagate:}\\
  &\bigwedge_{t\in V(T)}
  \bigwedge_{\alpha\assignment\bag(t)\cap Y}
  \Big[\,
    \sat^{\alpha}_{\leq t}\leftrightarrow\bigwedge_{\mathclap{c\in\baglabel(t)}}\sat_c^{\alpha}\wedge\bigwedge_{\mathclap{t'\in\children(t)}}\sat^{\alpha}_{<t,t'}
    \,\Big],\tag*{(3)}
\end{align*}
\begin{align*}
  &\comment{// Propagate satisfiability:}\\
  &\bigwedge_{t\in V(T)}
  \bigwedge_{\alpha\assignment\bag(t)\cap Y}
  \bigwedge_{t'\in\children(t)}      
  \Big[\,
    \sat^{\alpha}_{< t,t'}\leftrightarrow
    \bigwedge_{\mathclap{\substack{\beta\assignment\bag(t')\cap Y\\\beta\cap\lits(\bag(t))=\alpha\cap\lits(\bag(t'))}}}\sat^{\beta}_{\leq t'}
    \,\Big].\tag*{(4)}
\end{align*}
Observe that the constraints (1)--(4) contain no variable from $Y$ (we
removed them by locally speaking about $\alpha$) and, furthermore,
constraints (1), (3), and (4) are pure propagations, which leave no
degree of freedom on the auxiliary variables. Hence, models of these
constraint only have freedom in the variables in $X$ within constraint
(2). We are left with the task to count only models that actually
satisfy the input formula, which we achieve with:
\[
\bigvee_{\alpha\assignment\bag(\rootOf(T))\cap Y}\hspace{-1em}\sat^{\alpha}_{\leq\rootOf(T)}.\tag*{(5)}
\]

\begin{lemma~}\label{lemma:pmc-to-sharp}
  There is an algorithm that, given a \Lang{cnf}
  $\psi$, a set $X\subseteq\vars(\psi)$, and a width-$k$ tree
  decomposition of $G_\psi$, outputs in time $O^*(2^k)$ a \Lang{cnf} $\psi'$ and a
  width-$(12\cdot 2^k)$ tree decomposition of $G_{\psi'}$ such
  that the projected model count of $\psi$ on $X$ equals $\#(\psi')$.
\end{lemma~}
\begin{proof~}
  The bound on the treewidth and correctness of the propagation
  constraints follows analog to
  Lemma~\ref{lemma:qbf-eliminate-exists}.  We are left with the task
  to show that the number of models of the resulting formula equals
  the number of assignments of $X$ that can be extended to a
  satisfying assignment.  Formally, if $\xi\coloneq\xi(\psi,X)$ is the
  produced encoding for $\psi$ and $X\subseteq\vars(\psi)$, we need to
  show that:
  \begin{align*}
    &|\{\,
    \alpha\cap X\mid \text{$\alpha\assignment\vars(\psi)$ and $\psi|\alpha=\emptyset$}
    \,\}|\\
    =\,&|\{\,
    \beta\mid \text{$\beta\assignment\vars(\xi)$ and $\xi|\beta=\emptyset$}
    \,\}|.\tag*{($\star$)}
  \end{align*}    
  We prove by induction over $|X|$ with base case
  $X=\emptyset$. Then constraints (1)--(5) have no degree of
  freedom and we obtain either a (trivially) satisfiable formula or a
  contradiction, depending on whether or not the input was
  satisfiable. Assume $|X|>0$ and pick an arbitrary variable $x\in
  X$. We can rewrite the left side of $(\star)$ as:
  \begin{align*}
    &|\{\,
    \alpha\cap X\mid \text{$\alpha\assignment\vars(\psi)$ and $\psi|\alpha=\emptyset$}
    \,\}|=\\    
    &\quad\hspace{.33cm}|\{\,
    \alpha\cap (X\setminus\{x\})\mid \text{$\alpha\assignment\vars(\psi|x)$ and $(\psi|x)|\alpha=\emptyset$}
    \,\}|\\
    &\quad\hbox to 0pt{+\hss}\hspace{.33cm}
    |\{\,
    \alpha\cap (X\setminus\{x\})\mid\text{$\alpha\assignment\vars(\psi|\neg x)$ and $(\psi|\neg x)|\alpha=\emptyset$}
    \,\}|.
  \end{align*}
  Since $x$ is a free variable in $\xi$ (it appears in (2), where it is eventually used to satisfy $\sat^{\alpha}_c$), we can rewrite the right side of $(\star)$ as:
  \begin{align*}
    &|\{\,
    \beta\mid \text{$\beta\assignment\vars(\xi)$ and $\xi|\beta=\emptyset$}
    \,\}|=\\    
    &\quad\hspace{.33cm}|\{\,
    \beta \mid \text{$\beta\assignment\vars(\xi|x)$ and $(\xi|x)|\beta=\emptyset$}
    \,\}|\\
    &\quad\hbox to 0pt{+\hss}\hspace{.33cm}
    |\{\,
    \beta \mid\text{$\beta\assignment\vars(\xi|\neg x)$ and $(\xi|\neg x)|\beta=\emptyset$}
    \,\}|.
  \end{align*}
  From the induction hypothesis we obtain for both possible
  assignments of $x$:
  \begin{align*}
    &|\{\,
    \alpha\cap (X\setminus\{x\})\mid \text{$\alpha\assignment\vars(\psi|x)$ and $(\psi|x)|\alpha=\emptyset$}
    \,\}|\\
    &\quad= 
    |\{\,
    \beta\mid \text{$\beta\assignment\vars(\xi|x)$ and $(\xi|x)|\beta=\emptyset$}
    \,\}|;\\[1ex]
    &|\{\,
      \alpha\cap (X\setminus\{x\})\mid
      \text{$\alpha\assignment\vars(\psi|\neg x)$ and $(\psi|\neg x)|\alpha=\emptyset$}
      \,\}|\\
      &\quad = 
      |\{\,
      \beta\mid \text{$\beta\assignment\vars(\xi|\neg x)$ and
      $(\xi|\neg x)|\beta=\emptyset$}
      \,\}|.
  \end{align*}
  By substituting these back into $(\star)$, we conclude that
  $(\star)$ holds.
\end{proof~}

\begin{proof}[Proof of Theorem~\ref{theorem:pmctw}]
  By applying Fact~\ref{fact:semiringtw} to the formula generated by
  Lemma~\ref{lemma:pmc-to-sharp} we obtain an algorithm for \Lang{pmc}
  with running time $\tower(2, k + 3.59)$.
\end{proof}

%
%
\section{A SAT Version of Courcelle's Theorem}
\label{section:mso-to-sat}
\tcsautomoveaddto{main}{\subsection{Proofs for Section~\ref{section:mso-to-sat}}}

We demonstrate the power of treewidth-aware encodings
by providing an alternative proof of Courcelle's theorem. 
We prove the main part of Theorem~\ref{theorem:sattheorem} in the
following form:

\begin{lemma}\label{lemma:mctosat}
  There is an algorithm that, given a relational structure~$\structure$, a width-$k$ tree decomposition of $\structure$, and an
  \Lang{mso} sentence~$\phi$ in prenex normal form, produces in time
  $\tower(\qa(\phi),(k+9)|\phi|+3.92)$ a propositional formula
  $\psi$ and tree decomposition of $G_\psi$ of width $\toweronly(\qa(\phi),(k+9)|\phi|+3.92)$
  such that $\structure\models\phi\Leftrightarrow\psi\in\Lang{sat}$.
\end{lemma}

The lemma assumes that the sentence is in prenex normal form
with a quantifier-free part~$\psi$ in \Lang{cnf}, i.e.,
\(
   \phi\equiv Q_1 S_1\dots Q_{q-1} S_{q-1} Q_{q} s_q \dots Q_{\ell}s_\ell\suchthat\bigwedge_{i=1}^p\psi_i
\)
with $Q_i\in\{\exists,\forall\}$ and $S_i$ ($s_i$) being 
second-order (first-order) variables. The requirement that the second-order
quantifers appear before the first-order ones is for sake
of presentation, the encoding works as is if the 
quantifiers are mixed. The main part of the proof is a treewidth-aware
encoding from $\Lang{mc(mso)}$ into $\Lang{qsat}$;
which is then translated to $\Lang{sat}$ using
Theorem~\ref{theorem:qbftw}. 

\subsection{Auxiliary Encodings}
\label{sec:aux}

Let $\psi$ be a propositional formula and $X\subseteq \lits(\psi)$ be an arbitrary set
of literals. A \emph{cardinality constraint} $\card_{\bowtie c}(X)$
with $\bowtie\;\mathrel{\in}\{\leq,=,\geq\}$ ensures that \{ at most, exactly, at
least \} $c$ literals of $X$ get assigned to
true. Classic encodings of cardinality constraints increase the
treewidth of $\psi$ by quite a lot. For instance, the naive encoding
for $\card_{\leq 1}(X)\equiv\bigwedge_{\smash{u,v\in X; u\neq v}}(\neg u\vee\neg v)$
completes~$X$ into a clique. We encode a cardinality constraint
without increasing the treewidth by distributing a
\emph{sequential unary counter:}

\begin{lemma~}\label{lemma:cardinality}
  For every $c\geq 0$ we can, given a \Lang{cnf} $\psi$, a set
  $X\subseteq\lits(\psi)$, and a width-$k$ tree decomposition of
  $\psi$, encode $\card_{\bowtie c}(X)$ such that
  $\tw(\psi\wedge\card_{\bowtie c}(X))\leq k+3c+3$.
\end{lemma~}
\begin{proof~}
  \def\watch{\mathrm{watch}}\def\counter{\mathrm{counter}}%
  We define a
  mapping $\watch\colon V(T)\rightarrow(\lits(\phi)\cap X)\cup\{0\}$
  (0 refers to ``false'') with:
  \begin{enumerate}
  \item $\watch(t)=0$ if $|\children(t)|\neq 1$ for all $t\in V(T)$;
  \item for all $\ell\in X$ there is \emph{exactly one} $t\in V(T)$
    with $\watch(t)=\ell$;
  \item if $\watch(t)=\ell$, then the variable corresponding to $\ell$
    appears in $\bag(t)$.
  \end{enumerate}
  The existence of such a mapping can be ensured by copying bags of
  the given tree decomposition. We realize a sequential unary counter that we distribute along the
  tree decomposition and that we eventually increase at bags~$t$ with
  $\watch(t)\neq 0$. To realize a counter that can count up to~$c+1$,
  we add $c+1$ variables $\counter[t][1\dots c+1]$ to every bag $t\in V(T)$.
  The semantic is that $\counter[t][\cdot]$ is a unary counter that
  counts the literals in $X$ that are set to true in the subtree
  rooted at $t$. We may assume that the bags of leafs are empty and do not watch a
  variable. Hence, these counters need to be constant zero, which we enforce with:
  \begin{align*}
    \bigwedge_{\substack{t\in V(T)\\|\children(t)|=0}}
    \bigwedge_{i=1}^{c+1}
    \neg\counter[t][i].\tag*{(1)}
  \end{align*}
  For nodes $t\in V(T)$ that have one child, we have to propagate the
  counter of the child (recall that we count the literals in the
  subtree rooted at $t$) and, eventually, have to increase the counter
  if $\watch(t)$ is a true literal. The following two constraints
  realize this semantic:
  \begin{align*}
    &\bigwedge_{\substack{t\in V(T)\\\children(t)=\{t'\}}}\hspace{-2em}
    \counter[t][1]\leftrightarrow\counter[t'][1]\vee\watch(t);\tag*{(2)}\\
    &\bigwedge_{\substack{t\in V(T)\\\children(t)=\{t'\}}}
    \bigwedge_{i=2}^{c+1}\big(
    \counter[t][i]\leftrightarrow\counter[t'][i]\vee\big(\counter[t'][i-1]\wedge\watch(t)\big)\big).\tag*{(3)}
  \end{align*}
  Finally, we have to propagate the counter at decomposition nodes $t\in V(T)$ with
  two children, i.\,e., with $\children(t)=\{t',t''\}$. By the first property
  of the mapping we have $\watch(t)=0$
  (i.\,e., we do not need to worry about a local literal that we may
  need to add to the counter). Due to the second property, every literal in
  the subtree rooted at $t$ was counted at most once, i.\,e., the
  counter of $t$ has to be precisely the sum of the counters of $t'$
  and $t''$:
  \[
    \counter[t][\cdot] = \counter[t'][\cdot] + \counter[t''][\cdot].
  \]
  We can encode the addition of two unary counters in \Lang{cnf} with
  the following constraint:
    {\scriptsize\begin{align*}
      \bigwedge_{\substack{t\in V(T)\\\children(t)=\{t',t''\}}}
      \bigwedge_{i=1}^{c+1}
      \counter[t][i]
      \leftrightarrow
      \hspace{-2em}\bigvee_{\substack{a,b\in\{1,\dots,c+1\}\\ i=a+b}}\hspace{-2em}
      \big(
      \counter[t'][a] \wedge \counter[t''][b]
      \big)
      .\tag*{(4)}
  \end{align*}}
  The correctness of constraint (1)--(4) emirates from the
  aformentions ideas and the correctness of the sequential
  counter encoding~\cite{Sinz05}. We conclude that
  $\counter[\rootOf(T)][\cdot]$ stores the minimum of $c+1$ and the
  number of literals in $X$ that are set to true. In order to encode
   $\card_{\bowtie c}(X)$ for
  $\bowtie\;\in\{\leq,=,\geq\}$, we simply add the constraints
  $\neg\counter[\rootOf(T)][c+1]$;
  $\counter[\rootOf(T)][c]\wedge\neg\counter[\rootOf(T)][c+1]$; or
  $\counter[\rootOf(T)][c]$, respectively.

  Since we added $c+1$
  variables to every bag, clauses that
  only contain these variables and variables from the corresponding
  bag cannot increase the treewidth by more than $c+1$. To cover the propagation constraints (with (4) being the
  most involved one), we can virtually add to a bag the counter of its
  children. Since every node
  has at most two children, the treewidth increases overall by at most
  $3c+3$, that is, $\tw(\psi\wedge\card_{\bowtie c})\leq k+3c+3$
\end{proof~}

The second auxiliary encoding is a
treewidth-preserving conversion from \Lang{cnf}s to \Lang{dnf}s.

\begin{lemma~}\label{lemma:cnf-to-dnf}
  There is a polynomial-time algorithm that, given a \Lang{cnf}~$\psi$ and a width-$k$
  tree decomposition of $G_\psi$, produce a  
  \Lang{dnf}~$\psi'$ and a width-$(k+4)$ tree decomposition of $G_{\psi'}$
  such that for any $\alpha\sqsubseteq\var(\psi)$,
  $\psi|\alpha=\emptyset$ iff~$\psi'|\alpha$ is a tautology 
  ($\neg (\psi'|\alpha)$ is unsatisfiable). 
\end{lemma~}
\begin{proof~}
Let $(T,\bag,\baglabel)$ be the tree decomposition of $\psi$. For every clause
$C=(c_1\vee c_2\vee\dots\vee c_{|C|})$ of $\psi$ we introduce a
variable $f_C$ that indicates that $C$ is true: We need to negate the
constraint, as we are aiming for a tautology, i.\,e., intuitively, we
allow (skip) those assignments over variables~$f_C$ that invalidate
our constraints.
  \begin{align*}
    &\bigvee_{C\in\mathrm{clauses}(\psi)}
    \neg\Big[\,
    f_C \leftrightarrow (c_1\vee c_2\vee\dots\vee c_{|C|})
    \,\Big].\tag*{(1)}
  \end{align*}
  Note that this formula is a \Lang{dnf}. We encode satisfiability
  guided a long the tree decomposition by adding a variable $f_{\leq
    t}$ for every node $t\in V(T)$ that indicates that $\psi$ is
 satisfied in the subtree rooted at $t$. This semantic can be
 encoded with the following \Lang{dnf}:
  \begin{align*}
    &\bigvee_{t\in V(T)}
    \neg\Big[\,
    f_{\leq t}\,\leftrightarrow\hspace{-1em}\bigwedge_{C\in\baglabel(t)}\hspace{-1em}f_C\,\,\wedge\hspace{-1em}\bigwedge_{t'\in\children(t)}\hspace{-1em}f_{\leq t'}
      \,\Big].\tag*{(2)}
  \end{align*}%
  Again, the constraint we want to achieve appears negated.
  We conclude the construction by adding the term $f_{\leq\rootOf(T)}$ to~$\psi'$.
  Indeed, for any assignment~$\alpha\sqsubseteq \var(\psi)$,
  we have~$\psi|\alpha=\emptyset$ if and only if for any assignment~$\alpha'\sqsubseteq {\var(\psi')}$ with~$\alpha'\subseteq\alpha$ we have~$\emptyset \in \psi'|\alpha'$.
\end{proof~}

\subsection{Indicator Variables for the Quantifiers}\label{section:quantifers}
To prove Lemma~\ref{lemma:mctosat} we construct a \Lang{qbf}
for a given \Lang{mso} sentence $\phi$, structure $\structure$, and
tree decomposition of $\structure$. We first
define the primary variables of $\psi$, i.e., the prefix of $\psi$
(primary here refers to the fact that we will also need some auxiliary
variables later). For every second-order quantifier $\exists X$ or
$\forall X$ we introduce, as we did in the introduction, an indicator
variable $X_u$ for every element $u\in U(\structure)$ with the semantic that
$X_u$ is true iff $u\in X$. These variables are either existentially
or universally quantified, depending on the second-order
quantifier. If there are multiple quantifiers (say $\exists X\forall
Y$), the order in which the variables are
quantified is the same as the order of the second-order quantifiers.
For first-order quantifiers $\exists x$  or $\forall x$
we do the same construction,
i.e., we add variables~$x_u$ for all $u\in U(\structure)$
with the semantics that $x_u$ is true iff $x$ was assigned to $u$. Of
course, of these variables we have to set \emph{exactly one} to true,
which we enforce by adding 
$\card_{=1}(\{x_u\mid u\in U(\structure)\})$ using
Lemma~\ref{lemma:cardinality}. 

\subsection{Evaluation of Atoms}\label{sec:eval}
The last ingredient of our \Lang{qbf} encoding is the evaluation of
the atoms in the \Lang{mso} sentence $\phi$. An atom is
$Rx_1,\dots,x_a$ for a relational symbol $R$ from the vocabulary of arity
$a$, containment in a second-order variable $Xu$, equality $x=y$, and
the negation of the aforementioned. For every atom $\iota$ that
appears in $\phi$ we introduce variables $p^{\iota}_t$ and
$p^{\iota}_{\leq t}$ for all $t\in V(T)$ that indicate that $\iota$ is
true in bag $t$ or somewhere in the subtree rooted at $t$,
respectively. Note that the same atom can occur multiple times in
$\phi$, for instance in
\[
  \forall x\forall y\exists z\suchthat(x=y\rightarrow x=z)\vee(x=y\rightarrow y=z)
\]
there are two atoms $x=y$. However, since $\phi$ is in prenex normal
form (and, thus, variables cannot be rebound), these always evaluate in
exactly the same way. Hence, it is sufficient to consider the \emph{set
of atoms}, which we denote by~$\atoms(\phi)$.
We can propagate information about the atoms along the tree decomposition with:
\[
  \bigwedge_{t\in V(T)}\bigwedge_{\iota\in\atoms(\phi)} \Big[\,
  p^\iota_{\leq t}\leftrightarrow
  (p^{\iota}_t\,\vee\hspace{-1em}\bigvee_{t'\in\children(t)}\hspace{-1em}p^{\iota}_{\leq t'})
  \,\Big].
\]
This encoding introduces two variables per atom $\iota$ per bag $t$
(namely $p^{\iota}_t$ and $p^{\iota}_{\leq t}$), which increases the
treewidth by at most $2\cdot|\atoms(\phi)|$. 
To synchronize with the two children $t'$ and $t''$, we add $p^{\iota}_{\leq t'}$ 
and $p^{\iota}_{\leq t''}$ to $\bag(t)$, yielding a
total treewidth of at most $4\cdot|\atoms(\phi)|$. 

An easy atom to evaluate is $x=y$, since if $x$ and $y$ are equal
(i.e., they both got assigned to the same element $u\in U(\structure)$), we
can conclude this fact within a bag that contains $u$:
\[
  \bigwedge_{t\in V(T)} \Big[\,
  p^{x=y}_{t}\leftrightarrow
  \bigvee_{u\in\bag(t)}(x_u\wedge y_u)
  \,\Big].
  \]
For every $u\in U(\structure)$ and every quantifier $\exists x$ (or
$\forall x$), we add the propositional variable $x_u$ to all bags
containing $u$. We increase the treewidth by at most the quantifier
rank and, in return, cover constraints as the above trivially.
Similarly, if there is a second-order variable~$X$ and a first-order variable $x$, the atom
$Xx$ can be evaluated locally in every bag:
\[
  \bigwedge_{t\in V(T)} \Big[\,
  p^{X_x}_{t}\leftrightarrow
  \bigvee_{u\in\bag(t)}(X_u\wedge x_u)
  \,\Big].
\]
We have to evaluate atoms corresponding to relational symbols
$R$ of the vocabulary. For each such symbol of arity $a$ we encode:
\[
  \bigwedge_{t\in V(T)} \Big[\,
  p^{R(x_1,x_2,\dots,x_a)}_{t}\,\leftrightarrow
  \hspace{-1.5em}\bigvee_{\substack{u_1,\dots,u_a\in\bag(t)\\(u_1,\dots,u_a)\in R^{\structure}}}\hspace{-1.5em}
  \big((x_1)_{u_1}\wedge (x_2)_{u_2}\wedge\dots\wedge (x_a)_{u_a}\big)
  \,\Big].
\]
Here ``$R(x_1,x_2,\dots,x_a)$'' is an atom in which $R$ is a
relational symbol and $x_1,x_2,\dots,x_a$ are quantified first-order
variables. In the inner ``big-or'' we consider all 
$u_1,\dots,u_a$ in $\bag(t)$, i.e., elements $u_1,\dots,u_a\in U(\structure)$ that are in the relation
$(u_1,\dots,u_a)\in R^{\structure}$. Then ``$(x_i)_{u_i}$'' is a
variable that
describes that $x_i$ gets assigned to $u_i$. Note that all tuples in
$R^{\structure}$ appear together in at least one bag of the tree
decomposition and, hence, there is at least one bag $t$ for which
$p^{R(x_1,x_2,\dots,x_a)}_{t}$ can be evaluated to true. 
The propagation ensures that, for every 
$\iota\in\atoms(\phi)$, the variable $p^{\iota}_{\leq\rootOf(T)}$
will be true iff $\iota$ is true. Since the quantifier-free part of $\phi$ is a \Lang{cnf}
$\bigwedge_{j=1}^p\psi_j$, we can encode it by replacing every
occurrence of $\iota$ in $\psi_j$ with $p^{\iota}_{\leq\rootOf(T)}$.

\subsection{The Full Encoding in one Figure}
For the readers convenience, we compiled
the encoding into Figure~\ref{figt:redp}. Combining the insights of the last sections proves
Lemma~\ref{lemma:mctosat}, but if the inner-most quantifier is universal,
existentially projecting the encoding variables would produce a
\Lang{qbf} with one more block. This can, however, be circumvent using
Lemma~\ref{lemma:cnf-to-dnf}. We dedicate
Section~\ref{section:mso-to-sat-details} in the appendix to formally prove that
``combining the insights'' indeed leads to a sound proof of
Lemma~\ref{lemma:mctosat}.

\begin{figure}[htbp]
{
\begin{flalign}
& \textbf{Cardinality Propagation}\hspace{-40em}\notag\\
%
%
%
%
%
	%
	\label{redt:card}&c_{\leq t}^{x} \leftrightarrow\hspace{-2em} \bigvee_{u\in 
	\bag(t)\setminus\bag(\parent(t))} \hspace{-2em}x_u\hspace{0.5em}\vee\hspace{-0.5em} \bigvee_{t'\in\children(t)}\hspace{-1.5em}c_{\leq t'}^{x} \hspace{-15em}  & \text{for every }t\text{ in }T, x\in \{s_q, \ldots, s_\ell\}\\
& \textbf{At-Least-One Constraint}\hspace{-40em}\notag\\
	\label{redt:cardroot}&c_{\leq \rootOf(T)}^x\hspace{-15em} &\text{for every }x\in \{s_q, \ldots, s_\ell\}\\ 
& \textbf{At-Most-One Constraint}\hspace{-40em}\notag\\
	\label{redt:exclude}&\neg x_{u} \vee \neg x_{u'}\hspace{-15em} &\text{for every }t\text{ in }T, u,u'\in\bag(t), u\neq u', x\in \{s_q, \ldots, s_\ell\}\\
	%
	%
	%
	%
	\label{redt:choose}&\neg x_u \vee  \neg c_{\leq t'}^x\hspace{-15em}  & \text{for every }t\text{ in }T, t'\in\children(t), u\in\bag(t){\setminus}\bag(\parent(t)), x{\in} \{s_q, \ldots, s_\ell\}
	\\
	\label{redt:choosechld}&\neg c_{\leq t'}^x \vee  \neg c_{\leq t''}^x\hspace{-15em}  & \text{for every }t\text{ in }T, t', t''\in\children(t), 
	t'\neq t'', x\in \{s_q, \ldots, s_\ell\}\\
& \textbf{Proofs of MSO Atoms}\hspace{-40em}\notag\\
	%
	%
	%
	%
	\label{red:proveeq}&p_t^{x{=}y} \hspace{-.5em}\leftrightarrow \hspace{-1.15em}\bigvee_{u\in\bag(t)} \hspace{-.85em}(x_u \wedge y_u)\hspace{-15em}  & \text{for every }t\text{ in }T, 
	x,y\in\{s_q,\ldots,s_\ell\},
	(x{=}y){\in}\edges(\varphi)\\ 
	\label{red:prove}&p_t^{X(x)} \hspace{-.5em}\leftrightarrow \hspace{-1em}\bigvee_{u\in\bag(t)} \hspace{-0.75em}(X_u \wedge x_u)\hspace{-15em}  & \text{for every }t\text{ in }T, 
	X{\in}\{S_1,\ldots,S_{q-1}\},
	x{\in}\{s_q,\ldots,s_\ell\},
	X(x){\in}\edges(\varphi)\\
	%
	%
&p_t^{R(x_1,\ldots, x_a)} \leftrightarrow \hspace{-2em}\bigvee_{\substack{u_1,\dots,u_a\in\bag(t)\\(u_1,\dots,u_a)\in R^{\structure}}}\hspace{-2em} (
{(x_1)}_{u_1}\wedge \dots \wedge {(x_a)}_{u_a}) \hspace{-15em}  & \text{for every }t\text{ in }T, \{x_1,\ldots,x_a\}\subseteq \{s_q, \ldots, s_\ell\},
\notag\\[-1.5em]
\label{red:provee}&& R\in\structure, R(x_1,\ldots,x_r)\in \edges(\varphi)\\
	\label{red:provedown}&p_{\leq t}^{\iota} \leftrightarrow p_t^{\iota} \vee \hspace{-1.5em}\bigvee_{t'\in\children(t)} \hspace{-1.5em}p_{\leq t'}^{\iota}\hspace{-15em}  & \text{for every }t\text{ in }T, \iota\in \edges(\varphi)\\
	%
	%
	%
	%
& \textbf{Deriving MSO Atoms requires Proof}\hspace{-40em}\notag\\
	\label{red:provefin}&\iota \leftrightarrow p_{\leq \rootOf(T)}^{\iota}\hspace{-10em}  & \text{for every }t\text{ in }T, \iota\in \edges(\varphi)
	\\
	& \textbf{Verify MSO Formula}\hspace{-40em}\notag\\
	\label{red:form}&\psi & 
	%
\end{flalign}~\\[-2.75em]
}\caption{
The reduction~$\mathcal{R}_{\Lang{mso}\rightarrow \Lang{qsat}}(\phi,
\structure, \mathcal{T})$ that takes as input an \Lang{mso} formula in
prenex normal form~$\phi=Q_1 S_1\dots Q_{q-1} S_{q-1} Q_{q} s_q \dots Q_{\ell} s_\ell\suchthat \psi$ and a structure~$\structure$ with a TD $\mathcal{T}{=}(T,\bag)$ of~$\structure$ of width~$k$.
It obtains a QBF~$\phi'=Q_1 S_1'\dots Q_{\ell} S_\ell' \exists E' \suchthat \psi'$,
where $\psi'$ is the conjunction of Equations~(\ref{redt:card})--(\ref{red:form}), $S'_i=\{{(S_i)}_u \mid u\in U(\structure)\}$ and $E'=\var(\psi')\setminus(\bigcup_{i=1}^{\ell} S'_i)$. Formula~$\psi'$ can be easily converted into \Lang{cnf} of width linear in~$k$ (for constant-size \Lang{mso} formulas~$\phi$).}
\label{figt:redp}
\end{figure}

%
%
\section{Fagin Definability via Automated Reasoning}
\label{section:mso-optimization-and-counting}
\tcsautomoveaddto{main}{\subsection{Proofs for Section~\ref{section:mso-optimization-and-counting}}}

In this section we prove the remaining two items of
Theorem~\ref{theorem:sattheorem}, i.e., a treewidth-aware encoding
of the optimization version of Courcelle's Theorem to \Lang{maxsat};
and a \Lang{\#sat} encoding of the counting version of the
theorem. The general approach is as follows: We
obtain a \Lang{mso} formula $\phi(X)$ with a free set variable~$X$
as input (rather than a \Lang{mso} sentence as in
Lemma~\ref{lemma:mctosat}). The objective of the model-checking
problems adds requirements to this variable (for \Lang{fd(mso)} we
seek a $S\subseteq U(\structure)$ of minimum size such that
$\structure\models\phi(S)$; for $\#\Lang{fd(mso)}$ we want to count
the number of sets $S\subseteq U(\structure)$ with
$\structure\models\phi(S)$). The ``trick'' is to rewrite $\phi(X)=\xi$
as $\phi'=\exists X\xi$ and apply Lemma~\ref{lemma:mctosat} to $\phi'$
in order to obtain a propositional formula~$\psi$. Observe that the
quantifier alternation of $\phi'$ may be one larger than the one of
$\phi$.

\begin{lemma~}\label{lemma:fdtomaxsat}
  There is an algorithm that, given a structure
  $\structure$ with weights $w_i\colon U(\structure)\rightarrow\mathbb{Q}$ for
  $i\in\{1,\dots,\ell\}$, 
  a width-$k$ tree decomposition of $\structure$, and an \Lang{mso} formula
  $\phi(X_1,\dots,X_\ell)$ in prenex normal form, 
  produces in time
  $\tower(\qa(\phi)+1,(k+9)|\phi|+3.92)$ a \Lang{wcnf} $\psi$ 
  and a tree decomposition of width
  $\toweronly(\qa(\phi)+1,(k+9)|\phi|+3.92)$ of $G_\psi$
  such that the maximum
  weight of any model of $\psi$ equals the maximum value of
  $\sum_{i=1}^{\ell}\sum_{s\in S_i}w_i(s)$ under 
  $S_1,\dots,S_\ell\subseteq U(\structure)$ with
  $\structure\models\phi(S_1,\dots,S_{\ell})$.
\end{lemma~}
\begin{proof~}
  For each free set variable $X_i\in\{X_1,\dots,X_\ell\}$ we introduce
  indicator variables $(X_i)_u$ for all elements $u\in U(\structure)$ with
  the semantic that $(X_i)_u$ is true if $u$ gets selected to
  $X$~--~exactly as we did in Section~\ref{section:quantifers}. To
  encode the optimization (i.\,e., we seek an assignment that
  maximizes
  $\sum_{i=1}^{\ell}\sum_{u\in U(\structure)}(X_i)_u\cdot w_i(u)$), we
  add \emph{soft} clauses $(X_i)_u$ with weight
  $w\big((X_i)_u\big)=w_i(u)$. The remaining part of $\phi$ gets encoded
  exactly as in Lemma~\ref{lemma:mctosat} (all clauses of this
  encoding are \emph{hard}), whereby we treat the new indicator
  variables as the once for existential second-order
  quantifers. Therefore, the treewidth of the encoding may increase by
  one additional layer in the tower if $\phi$ starts with a universal
  quantifier (then there is a virtual alternation between the free
  variables and the first quantifier). 
\end{proof~}

\begin{lemma~}\label{lemma:sharpfdtosharpsat}
  There is an algorithm that, given a relational structure
  $\structure$, 
  a width-$k$ tree decomposition of $\structure$, and an \Lang{mso} formula
  $\phi(X_1,\dots,X_\ell)$ in prenex normal form, 
  produces in time
  $\tower(\qa(\phi)+1,(k+9)|\phi|+3.92)$ a \Lang{cnf} $\psi$ 
  and a tree decomposition of width
  $\toweronly(\qa(\phi)+1,(k+9)|\phi|+3.92)$ of $G_\psi$
  such that
  the number of models of $\psi$ equals the number of sets
  $S_1,\dots,S_\ell\subseteq U(\structure)$ with $\structure\models\phi(S_1,\dots,S_{\ell})$.
\end{lemma~}
\begin{proof~}
  We apply the same idea as in Lemma~\ref{lemma:fdtomaxsat} and
  introduce an indicator variable  $(X_i)_u$ for all
  $X_i\in\{X_1,\dots,X_\ell\}$ and elements $u\in
  U(\structure)$. Again, we also simply use Lemma~\ref{lemma:mctosat}
  to translate $\phi$ together with these indicator variables into a
  \Lang{cnf} $\psi$. To count the number of sets
  $S_1,\dots,S_\ell\in U(\structure)$ with
  $\structure\models\phi(S_1,\dots,S_\ell)$, we have to count the
  number of models of~$\psi$ projected to
  $\bigcup_{i=1}^\ell\bigcup_{u\in U(\structure)}(X_i)_u$, i.\,e., we
  encode the problem as instance of \Lang{pmc}. Note that, as
  in Lemma~\ref{lemma:fdtomaxsat}, this may increase the size of the
  tower by one layer compared to Lemma~\ref{lemma:mctosat} if $\phi$
  starts with a universal quantifier.
  
  We conclude the proof by reducing the projected model counting
  problem to $\Lang{\#sat}$ using Lemma~\ref{lemma:pmc-to-sharp}, which
  apparently increases the tower by one additional layer again. However, a
  closer look at the chain of reductions shows that this last step
  technically is not necessary: The reduction from \Lang{mc(mso)} to
  \Lang{sat} is via \Lang{qsat} and removes quantifiers one by
  one. Then the outer layer (the free variables, which increased the
  size of the tower) are the only remaining variables that are not
  deterministic, i.\,e., the reduction already produces an instance of
  $\Lang{\#sat}$ and, thus, the height of the tower increases only by one.
\end{proof~}

\section{Lower Bounds for the Encoding Size of Model Checking
  Problems}
\label{section:lowerbounds}
\tcsautomoveaddto{main}{\subsection{Proofs for Section~\ref{section:lowerbounds}}}

We companion our \Lang{sat} encodings for \Lang{mc(mso)} with lower
bounds on the achievable encoding size under $\Class{ETH}$. The first
lower bound (Theorem~\ref{theorem:encoding-lb}) is obtained by an encoding from $\Lang{qsat}$ into
$\Lang{mc(mso)}$ that implies that \Lang{sat}
encodings of \Lang{mc(mso)} lead to faster \Lang{qsat} algorithms.

\begin{lemma~}\label{lemma:mso-lower-bound}%
  There is a polynomial-time algorithm that, given a $\Lang{qsat}$
  sentence $\psi$, outputs a structure $\structure$ and an
  $\Lang{mso}$ sentence $\phi$ with
  $\tw(\structure)\leq\tw(\psi)+1$ and $\qa(\phi)\leq\qa(\psi)+2$ such
  that $\structure\models\phi$ iff $\psi$ evaluates to true.
\end{lemma~}
\begin{proof~}
  Let $\psi$ be a quantified propositional formula
  $\psi=Q_1x_1Q_2x_2\dots Q_nx_n\suchthat\gamma(x_1,\dots,x_n)$ in which
  $\gamma$ is a~\Lang{cnf} with clauses $C_1,\dots, C_m$. We partition
  the variables into maximal consecutive blocks with the same
  quantifier, that is, $B_1=\{x_1,x_2,\dots,x_{i_1}\}$ are all
  quantified in the same way, $B_2=\{x_{i_1+1},\dots,x_{i_2}\}$ as
  well, but with the opposite quantifier. The last block then is
  $B_{\qr(\phi)+1}=\{x_{i_{\qr(\phi)}+1},\dots,x_n\}$.  Let us
  construct the \emph{(signed) incidence graph} of $\psi$ as graph
  structure $\structure$ with universe $U(\structure)=\{x_1,\dots,x_n,C_1,\dots,C_m\}$
  and the relations:  
  \begin{align*}
    \mathrm{block_i}^\structure &= B_i \qquad\text{for all $i\in\{1,\dots,\qa(\psi)+1\}$},\\
    \mathrm{clause}^\structure  &= \{ C_1,\dots, C_m \},\\
    \mathrm{pos}^\structure &= \{\,(x, C)\mid \text{
    $x$ is a variable and $C$ a clause of $\psi$ and \phantom{$\neg$}$x\in C$ }\,\},\\
    \mathrm{neg}^\structure &= \{\,(x, C)\mid \text{
    $x$ is a variable and $C$ a clause of $\psi$ and $\neg x\in C$
      \,\}.}
  \end{align*}
  Consider for instance  the following fully quantified formula:
  \[
    \psi = \exists x_1\exists x_2\exists x_3
    \forall x_4\forall x_5
    \suchthat\,
    (x_1\vee x_2\vee\neg x_5)
    \wedge
    (x_3\vee x_4)
    \wedge
    (x_1\vee\neg x_4).
  \]
  In the example we have two blocks, namely $B_1=\{x_1,x_2,x_3\}$ and $B_2=\{x_4,x_5\}$. The
  incidence graph then is the following structure:
  
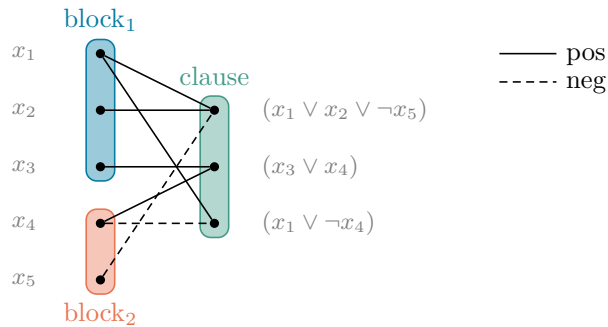
\begin{center}
    \begin{tikzpicture}[scale=0.65]
      \tikzset{
        dot/.style = {
          draw, fill, black, circle, inner sep=0pt, minimum width = 1mm,
          semithick
        },
        comment/.style = {gray, font=\small},
        edge/.style = {draw, semithick}
      }

      \node[dot] (x1) at (0,  0) {};
      \node[dot] (x2) at (0, -1) {};
      \node[dot] (x3) at (0, -2) {};

      \node[dot] (x4) at (0, -3) {};
      \node[dot] (x5) at (0, -4) {};

      \node[dot] (c1) at (2, -1) {};
      \node[dot] (c2) at (2, -2) {};
      \node[dot] (c3) at (2, -3) {};

      \draw[edge] (x1) to (c1);
      \draw[edge] (x1) to (c3);

      \draw[edge] (x2) to (c1);

      \draw[edge] (x3) to (c2);

      \draw[edge] (x4) to (c2);
      \draw[edge, densely dashed] (x4) to (c3);

      \draw[edge, densely dashed] (x5) to (c1);

      \foreach \i in {1,2,3,4,5}{
        \node[comment, left of=x\i] {$x_\i$};
      }
      \node[comment, right of=c1, anchor=west, xshift=-.5cm] {$(x_1\vee x_2\vee\neg x_5)$};
      \node[comment, right of=c2, anchor=west, xshift=-.5cm] {$(x_3\vee x_4)$};
      \node[comment, right of=c3, anchor=west, xshift=-.5cm] {$(x_1\vee\neg x_4)$};

      \begin{scope}[on background layer]
        \draw[rounded corners, semithick, color=fg, fill=fg!40, semithick] ($(x1)+(-.25,.25)$) rectangle ($(x3)+(.25,-.25)$);
        \node[color=fg, yshift=-.3cm] at ($(x1)+(0,1)$) {$\mathrm{block}_1$};

        \draw[rounded corners, semithick, color=bg, fill=bg!40, semithick] ($(x4)+(-.25,.25)$) rectangle ($(x5)+(.25,-.25)$);
        \node[color=bg, yshift=+.3cm] at ($(x5)-(0,1)$) {$\mathrm{block}_2$};

        \draw[rounded corners, semithick, color=gray, fill=gray!40, semithick] ($(c1)+(-.25,.25)$) rectangle ($(c3)+(.25,-.25)$);
        \node[color=gray, yshift=-.3cm] at ($(c1)+(0,1)$) {$\mathrm{clause}$};
      \end{scope}

      \draw[edge] (7,0)  -- ++(1, 0) node[right, yshift=-.25ex] {$\mathrm{pos}$};
      \draw[edge, densely dashed] (7,-.5) -- ++(1,0) node[right, yshift=-.25ex] {$\mathrm{neg}$};      
    \end{tikzpicture}
  \end{center}  
  To continue the proof, we consider the following \Lang{mso} sentence, where the $S_i$ are set variables:
  \begin{align*}
    \phi&\equiv Q_1S_1Q_{{i_1}+1}S_2\dots Q_{i_{\qa(\psi)}+1}S_{\qa(\psi)+1}
    \forall c\exists x
    \suchthat
    \mathrm{clause}(c)
    \rightarrow
    \bigvee_{j=1}^{\qa(\psi)+1}
    \big[
    \\    
    &\big(
    \mathrm{block}_j(x)
    \wedge
    \mathrm{pos}(x,c)
    \wedge
    S_j(x)
    \big)
    \vee
    \big(
    \mathrm{block}_j(x)
    \wedge
    \mathrm{neg}(x,c)
    \wedge
    \neg S_j(x)
    \big)\ \big].    
  \end{align*}
  An induction over $\qa(\psi)$ shows that $\structure$ is a
  satisfying assignment of $\phi$ iff $\psi$ evaluates to true. To
  conclude the proof, observe that the quantifier alternation in the
  second-order part of $\phi$ equals $\qa(\psi)$ and, hence,
  $\qa(\phi)\leq\qa(\psi)+2$ due to the $\forall x\exists y$
  block. Finally, we have $\tw(\structure)\leq\tw(\psi)+1$ since
  $\structure$ is the \emph{incidence} graph of $\psi$, whose
  treewidth is bounded by the \emph{primal} treewidth of $\psi$ plus one~[Chapter~17]\cite{HandbookSAT}.
\end{proof~}

\begin{proof}[Proof of Theorem~\ref{theorem:encoding-lb}]
  Combine Lemma~\ref{lemma:mso-lower-bound} with Fact~\ref{fact:qsat-lb}.
\end{proof}

\subsection{An Encoding for Compressing Treewidth}
\label{section:tw-saving}

For \Lang{qsat} one can ``move''
complexity from the quantifier rank of the formula to its
treewidth and \emph{vice versa}~\cite{FichteHP20}.  By
Lemma~\ref{lemma:mso-lower-bound}, this means that any
reduction from \Lang{qsat} to \Lang{mc(mso)} may produce an
instance with small treewidth or quantifier alternation while
increasing the other. We show that one can also
decrease the treewidth by increasing the \emph{block size}.

\begin{lemma~}\label{lemma:treewidthsaving}%
  For every $c>0$ there is a polynomial-time algorithm that, on input of a
  \Lang{cnf} $\psi$ and a width-$k$ tree decomposition of
  $G_\psi$, outputs a constant-size \Lang{mso} sentence $\phi$ with
  $\qa(\phi)=2$ and $\bs(\phi)=c$, and a
  structure $\structure$ with $\tw(\structure) \leq
  \lceil\frac{k+1}{c}\rceil$ such that
  $\psi\in\Lang{sat}\Leftrightarrow \structure\models\phi$.  
\end{lemma~}
\begin{proof~}
  The input of the theorem is a constant $c> 0$, a propositional formula
  $\psi$, and a width-$k$ tree decomposition $(T,\bag)$ of its primal
  graph. For a running example, consider
  \[
    \psi =
    \underbrace{(x_1\vee x_2\vee \neg x_5)}_{C_1}
    \wedge
    \underbrace{(\neg x_2\vee x_3\vee x_4\vee x_5)}_{C_2}
    \wedge
    \underbrace{(x_1\vee x_2\vee  \neg x_4\vee \neg x_5)}_{C_3},
  \]
  with its primal graph shown on the left and a corresponding tree
  decomposition on the right:

  \null\hfill
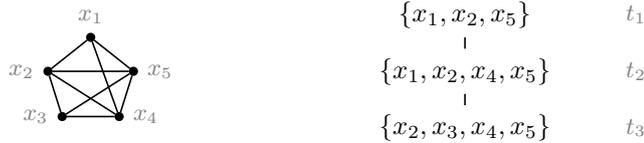
  \begin{tikzpicture}[scale=0.75, baseline=(x2)]
    \tikzset{
      dot/.style = {
        draw, fill, black, circle, inner sep=0pt, minimum width = 1mm,
        semithick
      },
      comment/.style = {gray, font=\small},
      edge/.style = {draw, semithick},
      every label/.style = {comment}
    }

    \node[dot, label=$x_1$]     (x1) at (0,-0.1)    {};
    \node[dot, label=180:$x_2$] (x2) at (-.75,-.7)  {};
    \node[dot, label=180:$x_3$] (x3) at (-.5,-1.5)  {};
    \node[dot, label=0:$x_4$]   (x4) at (.5,-1.5)   {};
    \node[dot, label=0:$x_5$]   (x5) at (.75,-.7)   {};
    
    \graph[use existing nodes, edges={edge}]{
      x2 -- {x3, x4, x5}; x3 -- {x4, x5}; x4 -- x5;
      x1 -- {x2, x4, x5}; 
    };
  \end{tikzpicture}
  \hfill
  \begin{tikzpicture}[scale=0.75, baseline=(2)]
    \tikzset{
      comment/.style = {gray, font=\small},
      edge/.style = {draw, semithick}
    }
    \node (1) at (0,0)  {$\{x_1, x_2, x_5\}$};
    \node (2) at (0,-1) {$\{x_1, x_2, x_4, x_5\}$};
    \node (3) at (0,-2) {$\{x_2, x_3, x_4, x_5\}$};
    \graph[use existing nodes, edges = {edge}]{
      1 -- 2 -- 3;
    };
    \foreach \b in {1,...,3}{
      \node[comment, right of = \b, xshift=1.25cm] {$t_\b$};
    }
  \end{tikzpicture}
  \hfill\null\smallskip
  
  We express the satisfiability problem as \Lang{mso}
  sentence similar to the one we used in the proof of
  Lemma~\ref{lemma:mso-lower-bound}. However, we add
  the available tree decomposition into the relational
  structure. Recall that we may assume that $(T,\bag)$ is a
  \emph{labeled tree decomposition} $(T,\bag, \baglabel)$ in which
  $\baglabel\colon V(T)\rightarrow\clauses(\psi)\cup\{\lambda\}$ assigns at most one
  clause of $\psi$ to any bag while every clause gets assigned to
  exactly one bag.
  The relational structure $\structure_1$ is
  an \emph{(signed) incidence graph} of $\psi$ on top of the given tree decomposition. Formally:
  \begin{align*}
    U(\structure_1) & =
    \{\,\langle v,t\rangle\mid \text{$v\in\bag(t)$ and $t\in V(T)$}\,\}\cup\clauses(\psi),\\
    \mathrm{clause}^{\structure_1} & = \clauses(\psi),\\
    \mathrm{pos}^{\structure_1} &=
    \{\, (C, \langle v,t\rangle) \mid \text{$\baglabel(t)=C$ and $\phantom{\neg}v\in C$} \,\},\\
    \mathrm{neg}^{\structure_1} &=
    \{\, (C, \langle v,t\rangle) \mid \text{$\baglabel(t)=C$ and $\neg v\in C$} \,\},\\
    \mathrm{sync}^{\structure_1} &=
    \{\, (\langle v,t\rangle,\langle v,t'\rangle) \mid \text{$v\in\bag(t)\cap\bag(t')$ and $(t,t')\in E(T)$} \,\}.    
  \end{align*}
  In words, these symbols mean that we have an element for every clause
  and for every variable per bag it appears in. The structure contains
  \emph{synchronization} edges between copies of the same variable; and
  we connected every clause to its variables in \emph{one} bag (either
  \emph{positively} or \emph{negatively}, depending of $\psi$). For
  our running example, $\structure_1$ would look like:
  \begin{center}
    \scalebox{0.75}
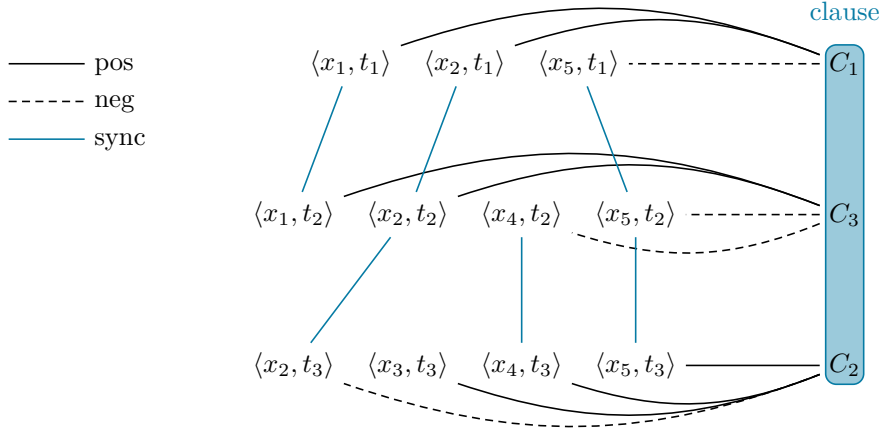
{\begin{tikzpicture}
      \tikzset{
        dot/.style = {
          draw, fill, black, circle, inner sep=0pt, minimum width = 1mm,
          semithick
        },
        comment/.style = {gray, font=\small},
        edge/.style = {draw, semithick},
        every label/.style = {comment}
      }
      
      \node (x1b1) at (-1.5,0)  {$\langle x_1,t_1\rangle$};
      \node (x2b1) at ( 0,0)    {$\langle x_2,t_1\rangle$};
      \node (x5b1) at (1.5,0)   {$\langle x_5,t_1\rangle$};

      \node (x1b2) at (-2.25,-2)  {$\langle x_1,t_2\rangle$};
      \node (x2b2) at (-.75,-2)   {$\langle x_2,t_2\rangle$};
      \node (x4b2) at (.75,-2)    {$\langle x_4,t_2\rangle$};
      \node (x5b2) at (2.25,-2)   {$\langle x_5,t_2\rangle$};

      \node (x2b3) at (-2.25,-4) {$\langle x_2,t_3\rangle$};
      \node (x3b3) at (-.75,-4)  {$\langle x_3,t_3\rangle$};
      \node (x4b3) at (.75,-4)   {$\langle x_4,t_3\rangle$};
      \node (x5b3) at (2.25,-4)  {$\langle x_5,t_3\rangle$};

      \node (c1) at (5,0) {$C_1$};
      \draw[edge] (c1)                 to[bend right=20] (x1b1);
      \draw[edge] (c1)                 to[bend right=20] (x2b1);
      \draw[edge, densely dashed] (c1) to                (x5b1);

      \node (c2) at (5,-4) {$C_2$};
      \draw[edge, densely dashed] (c2) to[bend left=20] (x2b3);
      \draw[edge]                 (c2) to[bend left=20] (x3b3);
      \draw[edge]                 (c2) to[bend left=20] (x4b3);
      \draw[edge]                 (c2) to               (x5b3);

      \node (c3) at (5,-2) {$C_3$};
      \draw[edge]                 (c3) to[bend right=20] (x1b2);
      \draw[edge]                 (c3) to[bend right=20] (x2b2);
      \draw[edge, densely dashed] (c3) to[bend left=20]  (x4b2);
      \draw[edge, densely dashed] (c3) to                (x5b2);

      \begin{scope}[on background layer]
        \draw[rounded corners, semithick, color=fg, fill=fg!40, semithick] ($(c1)+(-.25,.25)$) rectangle ($(c2)+(.25,-.25)$);
        \node[color=fg, yshift=-.3cm] at ($(c1)+(0,1)$) {$\mathrm{clause}$};
      \end{scope}

      \graph[use existing nodes, edges = {edge, fg}]{
        x1b1 -- x1b2;
        x2b1 -- x2b2 -- x2b3;
        x5b1 -- x5b2 -- x5b3;
        x4b2 -- x4b3;
      };

      \draw[edge] (-6,0)  -- ++(1, 0) node[right, yshift=-.25ex] {$\mathrm{pos}$};
      \draw[edge, densely dashed] (-6,-.5) -- ++(1,0) node[right, yshift=-.25ex] {$\mathrm{neg}$};      
      \draw[edge, draw=fg] (-6,-1) -- ++(1,0) node[right, yshift=-.25ex] {$\mathrm{sync}$};  
    \end{tikzpicture}}
  \end{center}
  Satisfiability of $\psi$ can be expressed with almost the same
  sentence as used in in the proof of
  Lemma~\ref{lemma:mso-lower-bound}, which was:
  \[\qquad\qquad
    \color{gray}
    \overbrace{\vphantom{\big(}\color{black}\exists
      S}^{\mathclap{\text{\it There is a set
          of variables s.t.}}}
    \quad
    \underbrace{\vphantom{\big(}\color{black}\forall x\exists y\suchthat
      \mathrm{clause}(x)\rightarrow}_{\mathclap{\text{\it for all clauses there is a variable}}}
    \quad
    \overbrace{\color{black}\big(\mathrm{pos}(x,y)\wedge
      Sy\big)}^{\mathclap{\text{\it appearing positive and being set to true}}}
    \quad
    \textcolor{black}{\vee}
    \quad
    \underbrace{\color{black}\big(\mathrm{neg}(x,y)\wedge \neg
      Sy\big)}_{\mathclap{\text{\it or negative and being set to
          false.}}}
    \!\textcolor{black}{.}
  \]
  The structure $\structure_1$ has the same information and
  relations as the structure constructed in the proof of
  Lemma~\ref{lemma:mso-lower-bound}, so the sentence is in
  principle still valid. However, since there are now copies of every
  variable, we have to ensure that either all copies of the same
  variable are quantified by $S$ or non of them is. We can achieve
  this with the synchronization edges:
  \begin{align*}
    \qquad\phi_1\mathrel{\equiv}\exists S\,&\hspace{1.6pt}\big[
    \forall x\exists y
    \suchthat
    \mathrm{clause}(x)\rightarrow
    \big(\mathrm{pos}(x,y)\wedge Sy\big)
    \vee
    \big(\mathrm{neg}(x,y)\wedge \neg Sy\big)
    \big]\mathrel{\wedge}\\
    &\color{gray}
    \underbrace{\color{black}\big[
    \forall x\forall y\suchthat
    \mathrm{sync}(x,y)
    \rightarrow
    \big(
    Sx\leftrightarrow Sy
      \big)\big]}_{\mathclap{\text{\it Variables connected by a
      sync edge can only appear together in $S$.}}}%
    \!\textcolor{black}{.}
  \end{align*}

  We continue by showing that this construction is correct and does
  not increase the treewidth (though, it also does not decrease it yet).

  \begin{claim}\label{claim:sat-mso}
    $\psi\in\Lang{sat}\Longleftrightarrow\structure_1\models\phi_1$
  \end{claim}
  \begin{proof}
    For the proof we ``move'' the second-order quantifier
    ($\exists S$) out of the formula and make it a free variable. For
    the first direction assume $\psi\in\Lang{sat}$ witnessed by an
    assignment $\beta\assignment\vars(\phi)$. Define
    $S=\beta\cap\vars(\phi)$, that is, $S$ is the set of variables
    appearing positively in $\beta$. Then, since $\beta\models\psi$,
    \emph{for all clause} $C\in\clauses(\psi)$ we either have $C\cap
    S\neq\emptyset$ (and, hence, \emph{there is a variable} appearing positively in
    $C$ that is in $S$), or $C\setminus S\neq\emptyset$ (thus, \emph{there
      is a variable} appearing negatively in $C$ that is not in
    $S$). By construction, all clauses appear in $\structure_1$ and are
    connected to the variables they contain, hence we have
    $\structure_1\models\phi_1(\, (S\times V(T))\cap U(\structure_1) \,)$.

    For the reverse direction first observe that, if there is a set
    $S\subseteq U(\structure_1)$ with
    $\structure_1\models\phi_1(S)$, then there is a set
    $S'\subseteq U(\structure_1)\cap(\vars(\psi)\times V(T))$ with
    $\structure_1\models\phi_1(S')$. This is because $\phi$ only
    speaks about $S$ via $y$, which is existentially bounded ($\exists
    y$) to be a neighbor of a clause:
    \[
      \color{gray}
     \textcolor{black}{ \mathrm{clause}(x)}
      \rightarrow
      (\textcolor{black}{\mathrm{pos}(x,y)}\wedge Sy)
      \vee
      (\textcolor{black}{\mathrm{neg}(x,y)}\wedge Sx).
    \]
    \noindent Observe that, due to
    $\forall x\forall y\suchthat \mathrm{sync}(x,y) \rightarrow \big(
    Sx\leftrightarrow Sy \big)$, we have for all $x_i\in\vars(\psi)$ either
    \begin{align*}
      &(\{x_i\}\times V(T))\cap U(\structure_1)\subseteq S'\\
      \quad\text{or}\quad&
      (\{x_i\}\times V(T))\cap U(\structure_1)\cap S'=\emptyset
      .
    \end{align*}
    Hence, $S'$ is uniquely determined by a set
    $S''\subseteq\vars(\psi)$, which can be extended to an
    assignment $\beta\assignment\vars(\psi)$ by adding all missing
    variables negated. Therefore $\psi\in\Lang{sat}$.
  \end{proof}
  \begin{claim}\label{claim:tw-bounded}
    $\tw(\structure_1)\leq k+1$
  \end{claim}
  \begin{proof}[Proof of Claim~\ref{claim:tw-bounded}]
    We may assume that in $(T,\bag)$ (the width-$k$ tree decomposition
    of~$\psi$) all bags are completed to a clique (this does
    not increase the treewidth since $(T,\bag)$ is still a valid tree
    decomposition). Then in $\structure_1$ all clauses $C$ are
    connected to a clique (recall that each clause is only connected
    to one bag) and, hence, they are \emph{simplicial}. By the simplicial
    rule~\cite{BodlaenderK06} the treewidth of $\structure_1$ is the maximum of the degree of $C$ (which
    is $|C|\leq k+1$) and the treewidth of the graph obtained by
    removing $C$. Hence, if $\structure_1'$ is obtained from
    $\structure_1$ by deleting all clauses, we have $\tw(\structure_1)\leq\max\{\,k+1,\tw(\structure_1')\,\}$.

    We are left with the task of showing
    $\tw(\structure_1')\leq k+1$. Observe that, if we ignore the
    sync edges, a tree decomposition of $\structure_1'$ is again
    a tree decomposition of $\psi$. To repair Property~2 of tree decompositions if
    the sync edges are present, we need to solve the following
    problem: Given two adjacent bags, say  $t_1=\{x_1,\dots,x_k\}$ and $t_2=\{y_1,\dots,y_k'\}$, of the tree decomposition with
    disjoint set of vertices that are connected by a matching, adapt
    the decomposition such that all edges of the matching are
    covered. For simplicity, we may assume $k=k'$ and that
    there is a perfect matching connecting $x_i$ with $y_i$. We inject
    a path in the tree decomposition between $t_1$ and $t_2$ of the
    following form:
    {\small\begin{align*}
      \{x_1,\dots,x_k\}
      &- \{x_1,\dots,x_k\}\setminus\{\}\cup\{y_1\}\\
      &- \{x_1,\dots,x_k\}\setminus\{x_1\}\cup\{y_1,y_2\}\\
      &- \{x_1,\dots,x_k\}\setminus\{x_1,x_2\}\cup\{y_1,y_2,y_3\}\\
      &- \dots\\
      &- \{x_1,\dots,x_k\}\setminus\{x_1,x_2,x_3,\dots,x_{k-1}\}\\&\qquad\cup\{y_1,y_2,y_3,\dots,y_k\}\\
      &- \{y_1,y_2,y_3,\dots,y_k\}.
    \end{align*}}
    Clearly, this increases the maximum bag size by at most one and
    restores all properties of a tree decomposition.
  \end{proof}
  To recap, we have now translated $\psi$ (and a width-$k$ tree
  decomposition of it) into an (constant size) \Lang{mso} sentence
  $\phi_1$ with $\qa(\phi_1)=2$ and a structure $\structure_1$ of treewidth
  $\tw(\structure_1)\leq k+1$ such that
  $\psi\in\Lang{sat}\Leftrightarrow\structure_1\models\phi_1$. 

  Of course, the goal of this prove is to construct a structure
  $\structure_2$ with a \emph{smaller} treewidth. The
  main idea to obtain $\structure_2$ from $\structure_1$ is
  to \emph{contract} vertices within a bag, i.\,e., vertices that
  correspond to variables and that appear together in a bag. More
  precisely, we partition every bag (of the given tree
  decomposition $(T,\bag)$) into $\lceil(k+1)/c\rceil$ groups
  that we then replace by a super vertex in~$\structure_2$.
  For instance, if $t\in V(T)$ is a bag of the
  original tree decomposition and we have vertices
  $\langle x_1,t\rangle,\dots,\langle x_k,t\rangle$ in
  $U(\structure_1)$, we would add
  $\big\langle\langle x_1,t\rangle,\dots,\langle
  x_c,t\rangle\big\rangle$,
  $\big\langle\langle x_{c+1},t\rangle,\dots,\langle
  x_{2c},t\rangle\big\rangle$ up to
  $\big\langle\langle x_{k-c},t\rangle,\dots,\langle
  x_k,t\rangle\big\rangle$ to $U(\structure_2)$. Note that the
  $x_i$ all correspond to variables, we do \emph{not} contract
  vertices corresponding to clauses. 

  We introduce the following relations to $\structure_2$: There
  are $c$ relations $\mathrm{pos}_i^2$ and $\mathrm{neg}_i^2$ for
  $c\in\{1,\dots,c\}$, respectively. The semantics of these is that
  $\mathrm{pos}_i^{\structure_2}(x,y)$ means that the clause~$x$
  contains the $i$th component of $y$ (which is a variable) positively
  (respectively negatively for
  $\mathrm{neg}_i^{\structure_2}$). Note that we use an order on
  the contracted vertices here (i.\,e., we refer to the ``$i$th
  component'' of a super vertex); we may do so by fixing (during the
  reduction) any order on the nodes of a bag. However, this order is
  later not revealed to the structure and only relevant for the
  definition of the relations. Similarly, we introduce $c^2$ relations
  $\mathrm{sync}_{i,j}^{2}$ for $i,j\in\{1,\dots,c\}$ that indicate
  that the $i$th component of the first argument has to be synced with
  the $j$th component of the second argument.
  Formally we define $\structure_2$ for all $i,j\in\{1,\dots,c\}$ as:
  \begin{align*}
    U(\structure_2) & =
                           \bigcup_{\mathclap{\stackrel{t\in V(T)}{\bag(t)=\{x_1,\dots,x_k\}}}}\,\{\,
                           \big\langle\langle x_1,t\rangle,\dots,\langle x_c,t\rangle\big\rangle,
                           \dots
                           \big\langle\langle x_{k-c},t\rangle,\dots,\langle x_k,t\rangle\big\rangle
                           \,\}\cup\clauses(\psi),
    \\
    \mathrm{clause}^{\structure_2} & = \clauses(\psi),\\
    \mathrm{pos}_i^{\structure_2} &=
    \{\, \big(C,  \big\langle\langle x_{\ell_1},t\rangle,\dots,\langle x_{\ell_i},t\rangle,\dots,\langle x_{\ell_c},t\rangle\big\rangle\big) \mid \text{$\pi(C)=t$ and $\phantom{\neg}x_{\ell_i}\in C$} \,\},\\
    \mathrm{neg}_i^{\structure_2} &=
    \{\, \big(C,  \big\langle\langle x_{\ell_1},t\rangle,\dots,\langle x_{\ell_i},t\rangle,\dots,\langle x_{\ell_c},t\rangle\big\rangle\big) \mid \text{$\pi(C)=t$ and $\neg x_{\ell_i}\in C$} \,\},\\
    \mathrm{sync}_{i,j}^{\structure_2} &=
    \{\, \big(\big\langle\langle x_{\ell_1},t\rangle,\dots,\langle x_{\ell_i},t\rangle,\dots,\langle x_{\ell_c},t\rangle\big\rangle,\big\langle\langle y_{\ell'_1},t\rangle,\dots,\langle y_{\ell'_j},t\rangle,\dots,\langle y_{\ell'_c},t\rangle\big\rangle\big) \mid \\
    &\quad\qquad \mathrm{sync}^{\structure}(\langle x_{\ell_i},t\rangle, \langle y_{\ell'_j},t'\rangle) \,\}.
  \end{align*}
  For the running example, the new structure would look as follows for
  $c=2$:
  \begin{center}
    \scalebox{0.75}
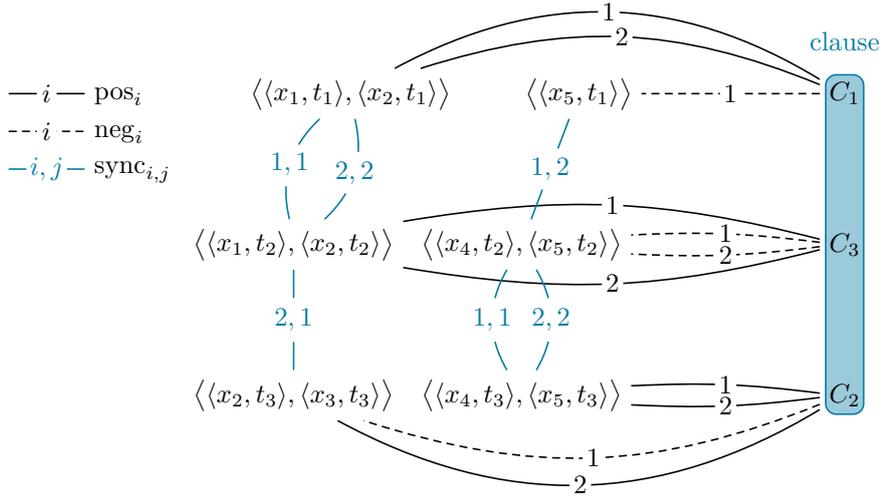
{\begin{tikzpicture}
      \tikzset{
        dot/.style = {
          draw, fill, black, circle, inner sep=0pt, minimum width = 1mm,
          semithick
        },
        comment/.style = {gray, font=\small},
        edge/.style = {draw, semithick},
        every label/.style = {comment}
      }
      
      \node (ab1) at (-1.5,0)  {$\big\langle\langle x_1,t_1\rangle,\langle x_2,t_1\rangle\big\rangle$};
      \node (bb1) at (1.5,0)   {$\big\langle\langle x_5,t_1\rangle\big\rangle$};

      \node (ab2) at (-2.25,-2)  {$\big\langle\langle x_1,t_2\rangle,\langle x_2,t_2\rangle\big\rangle$};
      \node (bb2) at (.75,-2)    {$\big\langle\langle x_4,t_2\rangle,\langle x_5,t_2\rangle\big\rangle$};

      \node (ab3) at (-2.25,-4) {$\big\langle\langle x_2,t_3\rangle,\langle x_3,t_3\rangle\big\rangle$};
      \node (bb3) at (.75,-4)   {$\big\langle\langle x_4,t_3\rangle,\langle x_5,t_3\rangle\big\rangle$};

      \node (c1) at (5,0) {$C_1$};
      \draw[edge] (c1)                 to[bend right=20] node[midway, fill=white, circle, inner sep=0pt] {$2$} (ab1);
      \draw[edge] (c1)                 to[bend right=30] node[midway, fill=white, circle, inner sep=0pt] {$1$} (ab1);
      \draw[edge, densely dashed] (c1) to                node[midway, fill=white, circle, inner sep=0pt] {$1$} (bb1);

      \node (c2) at (5,-4) {$C_2$};
      \draw[edge, densely dashed] (c2) to[bend left=20] node[midway, fill=white, circle, inner sep=0pt] {$1$} (ab3);
      \draw[edge]                 (c2) to[bend left=30] node[midway, fill=white, circle, inner sep=0pt] {$2$} (ab3);
      \draw[edge]                 (c2) to[bend left=5]  node[midway, fill=white, circle, inner sep=0pt] {$2$} (bb3);
      \draw[edge]                 (c2) to[bend right=5] node[midway, fill=white, circle, inner sep=0pt] {$1$} (bb3);

      \node (c3) at (5,-2) {$C_3$};
      \draw[edge]                 (c3) to[bend right=12] node[midway, fill=white, circle, inner sep=0pt] {$1$} (ab2);
      \draw[edge]                 (c3) to[bend left=12]  node[midway, fill=white, circle, inner sep=0pt] {$2$} (ab2);
      \draw[edge, densely dashed] (c3) to[bend left=5]   node[midway, fill=white, circle, inner sep=0pt] {$2$} (bb2);
      \draw[edge, densely dashed] (c3) to[bend right=5]  node[midway, fill=white, circle, inner sep=0pt] {$1$} (bb2);

      \begin{scope}[on background layer]
        \draw[rounded corners, semithick, color=fg, fill=fg!40, semithick] ($(c1)+(-.25,.25)$) rectangle ($(c2)+(.25,-.25)$);
        \node[color=fg, yshift=-.3cm] at ($(c1)+(0,1)$) {$\mathrm{clause}$};
      \end{scope}

      \draw[edge, fg] (ab1) to[bend right] node[midway, fill=white, circle, inner sep=0pt] {$1,1$} (ab2);
      \draw[edge, fg] (ab1) to[bend left]  node[midway, fill=white, circle, inner sep=0pt] {$2,2$} (ab2);
      \draw[edge, fg] (bb1) to node[midway, fill=white, circle, inner sep=0pt] {$1,2$} (bb2);
      \draw[edge, fg] (ab2) to node[midway, fill=white, circle, inner sep=0pt] {$2,1$} (ab3);
      \draw[edge, fg] (bb2) to[bend right] node[midway, fill=white, circle, inner sep=0pt] {$1,1$} (bb3);
      \draw[edge, fg] (bb2) to[bend left]  node[midway, fill=white, circle, inner sep=0pt] {$2,2$} (bb3);

      \draw[edge] (-6,0)  --node[midway, fill=white, circle, inner sep=0pt] {$i$} ++(1, 0) node[right, yshift=-.25ex] {$\mathrm{pos}_i$};
      \draw[edge, densely dashed] (-6,-.5) --node[midway, fill=white, circle, inner sep=0pt] {$i$} ++(1,0) node[right, yshift=-.25ex] {$\mathrm{neg}_i$};      
      \draw[edge, draw=fg] (-6,-1) --node[midway, fill=white, circle, inner sep=0pt] {\color{fg}$i,j$} ++(1,0) node[right, yshift=-.25ex] {$\mathrm{sync}_{i,j}$};  
    \end{tikzpicture}}
  \end{center}

  We can update $\phi_1$ to speak about $\structure_2$ as
  follows: We replace $\exists S$ with $c$ new second-order
  quantifiers, i.\,e., $\exists S_1\exists S_2\dots\exists S_c$. The
  semantics shall be that $S_i(x)$ means that the $i$th component of
  $x$ is in $S$. Formally we write:

  \begin{align*}
    \phi_2\equiv \exists S_1\exists S_2\dots\exists S_c\suchthat
    &\big[
      \forall x\exists y\suchthat
    \mathrm{clause}(x)\rightarrow
    \bigvee_{\mathclap{i=1}}^c
    \big(\mathrm{pos}_i(x,y)\wedge S_i(y)\big)
    \vee
    \big(\mathrm{neg}_i(x,y)\wedge \neg S_i(y)\big)
      \big]\\
    \wedge\,&\big[
      \forall x\forall y\suchthat
      \bigwedge_{i=1}^c\bigwedge_{j=1}^c \mathrm{sync}_{i,j}(x,y)\rightarrow(S_i(x)\leftrightarrow S_j(y))
      \big].
  \end{align*}

  \begin{claim}
    $\psi\in\Lang{sat}\Leftrightarrow\phi_2\models\structure_2$
  \end{claim}
  \begin{proof}
    We closely follow the proof of Claim~\ref{claim:sat-mso}. In fact,
    the proof is almost identical, the differences are:
    \begin{description}
      \item[In the first direction] we again consider a model
        $\beta$ of $\psi$ and define $S=\beta\cap\vars(\phi)$ to be
        the set of variables that are set to true. We have to
        ``distribute'' $S$ to the $S_i$ according to the way we
        contracted variables, i.\,e., we set
        \[
          S_i = \big\{\,
          \big\langle
          \langle x_{\ell_1},t\rangle,
          \dots,
          \langle x_{\ell_c},t\rangle
          \big\rangle
          \mid
          \text{$x_{\ell_i}\in S$ and $t\in V(T)$}
          \,\big\}.
        \]
        Hence $\structure_2\models\phi_2(S_1,\dots,S_c)$.
      \item[In the second direction] we can construct a model for
        $\psi$ by initializing $S=\emptyset$. Then we iterate over
        $i\in\{1,\dots,c\}$ and over all elements $e\in S_i$. For each
        such element, we add the $i$th component of $e$ to $S$. Finally, we
        construct a model $\beta$ of $\psi$ from $S$ by adding all
        variables that do not appear in $S$ negated to it.\qedhere
    \end{description}
  \end{proof}
  \begin{claim}
    $\tw(\structure_2)\leq \lceil(k+1)/c\rceil$
  \end{claim}
  \begin{proof}
    A tree decomposition of the claimed width can be obtained from a
    width-$k$ tree decomposition of $\structure_1$ by performing
    the contractions within the bags, since multi-edges do not
    increase the treewidth.
  \end{proof}
This completes the proof as $\phi_2$ has constant size (depending only
on $c$), quantifier alternation $2$, block size $c$, and since $\tw(\structure_2)\leq \lceil(k+1)/c\rceil$.
\end{proof~}

\begin{proof}[Proof of Theorem~\ref{theorem:tradeoff}]
  We obtain the Trade-off Theorem by combining the proof strategy of
  Lemma~\ref{lemma:treewidthsaving} with the reduction from \Lang{qsat}
  to \Lang{mc(mso)} of Lemma~\ref{lemma:mso-lower-bound}. The result
  is a polynomial-time algorithm for every $c>0$ that, on input of a
  \Lang{qbf} $\psi$ and a width-$k$ tree decomposition of
  $G_\psi$, outputs a constant-size \Lang{mso} sentence $\phi$ with
  $\qa(\phi)\leq\qa(\psi)+2$ and $\bs(\phi)=c$, and a
  structure $\structure$ with $\tw(\structure) \leq
  \lceil\frac{k+1}{c}\rceil$ such that
  $\psi$ is valid iff $\structure\models\phi$.  
\end{proof}

It is out of the scope of this article, but worth mentioning, that the
proofs of Lemma~\ref{lemma:treewidthsaving} and
Theorem~\ref{theorem:tradeoff} can be generalized to the following
finite-model theoretic result:

\begin{proposition}
  For every $c>0$ there is a polynomial-time algorithm that, given a relational
  structure $\structure$, a width-$k$ tree decomposition of $\structure$, and an \Lang{mso} sentence $\phi$, outputs a
  structure $\structure'$ and a sentence $\phi'$ such that:
  \begin{enumerate}
  \item $\structure\models\phi\Longleftrightarrow\structure'\models\phi'$;
  \item $\tw(\structure')\leq \lceil\frac{k+1}{c}\rceil$;
  \item $\bs(\phi')\leq c\cdot\bs(\phi)$.
  \end{enumerate}
\end{proposition}

%
%
\section{Conclusion and Further Research}
\label{section:conclusion}

We studied \emph{structure-guided automated reasoning}, where we
utilize the input's structure in propositional encodings. The
scientific question we asked was whether we can encode every
\Lang{mso} definable problem on structures of bounded treewidth into
\Lang{sat} formulas of bounded treewidth. We proved this in the
affirmative, implying an alternative proof of Courcelle's Theorem. The
most valuable aspects are, in our opinion, the simplicity of the proof
(it is ``just'' an encoding into propositional logic) and the
potential advantages in practice for formulas of small quantifier
alternation (\Lang{sat} solvers are known to perform well on instances
of small treewidth, even if they do not actively apply techniques such
as dynamic programming). Another advantage is the surprisingly simple
generalization to the optimization and counting version of Courcelle's
Theorem~--~we can directly ``plug in'' \Lang{maxsat} or \Lang{\#sat}
and obtain the corresponding results. As a byproduct, we also obtain
new proofs showing (purely as encodings into propositional logic) that
\Lang{qsat} parameterized by the input's treewidth plus quantifier
alternation is fixed-parameter tractable (improving a complex dynamic
program with nested tables) and that \Lang{pmc} parameterized by
treewidth is fixed-parameter tractable (improving a multi-pass dynamic
program). Table~\ref{table:overview} provides an overview of the
encodings presented within this article.

\begin{table}[htbp]
  \caption{We summarize the encodings presented within this
    article. An encoding maps \emph{from} one problem
    \emph{to} another. The third and fourth
    columns define the treewidth and size of the encoding, whereby we
    assume that $c>0$ is a constant, a width-$k$ tree decomposition is
    given, $\psi$ is a propositional formula, and $\phi$ is a
    fixed \Lang{mso} formula.}
  \label{table:overview}
  \resizebox{\textwidth}{!}{\begin{tabular}{lllll}
  \toprule
    \emph{Encoding\dots}        &                &                                                 &                                                  &                                         \\
    \emph{From}                 & \emph{To}      & \emph{Treewidth}                                & \emph{Size}                            
                                   & \emph{Reference}                                                                                                                               \\[2ex]
     \midrule
    \Lang{qsat}                 & \Lang{sat}     & $\toweronly(\qa(\psi),k+3.92)$                  & $\tower(\qa(\psi),k+3.92)$                 & Theorem~\ref{theorem:qbftw}       \\
    \Lang{pmc}                  & $\Lang{\#sat}$ & $\toweronly(1,k+3.59)$                          & $\tower(1,k+3.59)$                         & Theorem~\ref{theorem:pmctw}        \\[1ex]
    $\card_{\bowtie c}(X)$      & \Lang{sat}             & $k+3c+3$                                        & $O(c|\psi|)$                                      & Lemma~\ref{lemma:cardinality}           \\
    \Lang{cnf}                  & \Lang{dnf}     & $k+4$                                           & $O(|\psi|)$                                     & Lemma~\ref{lemma:cnf-to-dnf}            \\[1ex]
    \Lang{mc(mso)}              & \Lang{sat}     & $\toweronly(\qa(\phi),(9k+9)|\phi|+3.92)$       & $\tower(\qa(\phi),(9k+9)|\phi|+3.92)$             & Lemma~\ref{lemma:mctosat}               \\
    \Lang{fd(mso)}              & \Lang{maxsat}  & $\toweronly(\qa(\phi)+1,(9k+9)|\phi|+3.92)$   & $\tower(\qa(\phi)+1,(9k+9)|\phi|+3.92)$           & Lemma~\ref{lemma:fdtomaxsat}            \\
    $\Lang{\#fd(mso)}$          & $\Lang{\#sat}$ & $\toweronly(\qa(\phi)+1,(9k+9)|\phi|+3.92)$     & $\tower(\qa(\phi)+1,(9k+9)|\phi|+3.92)$           & Lemma~\ref{lemma:sharpfdtosharpsat}     \\[1ex]
    \Lang{sat}                  & \Lang{mc(mso)} & $\lceil\frac{k+1}{c}\rceil$                     & $O(k|\psi|)$                                      & Lemma~\ref{lemma:treewidthsaving}       \\
    \Lang{qsat}                 & \Lang{mc(mso)} & $\lceil\frac{k+1}{c}\rceil$                     & $O(k|\psi|)$                                      & Theorem~\ref{theorem:tradeoff}   \\\bottomrule
  \end{tabular}}
\end{table}

Our encodings are exponentially smaller than the best known running
time for $\Lang{mc(mso)}$, i.e., when we solve the instances using
Fact~\ref{fact:semiringtw}, we obtain the same runtime. We
complemented this finding with new $\Class{ETH}$-based lower
bounds. Further research will be concerned with closing the remaining
gap in the height of the tower between the lower and upper bounds. We
show in an upcoming paper that the terms ``$\qa(\phi)$'' in
Theorem~\ref{theorem:sattheorem} and ``$\qa(\phi)-2$'' in
Theorem~\ref{theorem:encoding-lb} can be replaced by ``$\qa_2(\phi)$''
on \emph{guarded} formulas, i.e., formulas in which there are only two
first-order quantifers that are only allowed to quantify edges. Here,
$\qa_2(\phi)$ refers to the quantifier alternation of the second-order
quantifers only. Hence, on such guarded formulas (e.g., on all
examples in the introduction), the bounds are tight. Another task that
remains for further research is to evaluate the encodings in
practice. This would also be interesting for the auxiliary encodings,
e.g., can a treewidth-aware cardinality constraint compete with
classical cardinality constraint?

\clearpage
\bibliography{main}

\tcsautomoveinsert{main}

\appendix
\section{Technical Details: Treewidth-Aware Encoding from MSO to QSAT}\label{section:mso-to-sat-details}
In this section we show how the
ideas of Section~\ref{section:mso-to-sat} can be glued together and prove the
correctness of the encoding.
Let~\[\phi=Q_1 S_1\dots Q_{q-1} S_{q-1} Q_{q} s_q \dots
Q_{\ell} s_\ell\suchthat \psi\] be an \Lang{mso} sentence, and suppose
we are given a structure~$\structure$ and a corresponding decomposition~$(T,\bag)$.  Technically, we describe a reduction
$\mathcal{R}_{\Lang{mso}\rightarrow\Lang{qsat}}$ that transforms~$\phi$ into
a \Lang{qbf} of the form $\varphi'=Q_1
S'_1\dots Q_{\ell} S'_\ell \exists E'\suchthat \psi'$, where sets $S'_i$ of
variables for~$1\leq i\leq q-0$ and~$q\leq j\leq \ell$ comprise
indicator variables, i.\,e.,
$S'_i=\{{(S_i)}_u \mid u\in U(\structure)\}$ and
$S'_j=\{{(s_j)}_u \mid u\in U(\structure)\}$, respectively.  Remaining
auxiliary variables are given by
$E'=\var(\psi')\setminus\bigcup_{i=1}^{\ell}(S'_i) $.

The figure contains an implementation of the cardinality constraint
$\card_{\bowtie c}(X)$ from Section~\ref{sec:aux} for the special case
$\card_{=1}(X)$, which is the case we need to encode first-order
variables (Equations~(\ref{redt:card})--(\ref{redt:choosechld})).  We
use auxiliary variables~$c_{\leq t}^x$ to guide the
information of whether~$x$ has been assigned an element from the
universe up to a node~$t$ of~$T$ along the tree.

Equations~(\ref{red:proveeq})--(\ref{red:provefin}) depict the
evaluation of atoms, where $p_t^\iota$ indicates
whether we find a witness for an atom~$\iota$. This is carried out by
Equations~(\ref{red:proveeq}), (\ref{red:prove}), and
(\ref{red:provee}) for equality expressions, set expressions, and
relational expressions, respectively.  The evaluation is guided along
the tree by means of Equation~(\ref{red:provedown}).  In the end, the
reduction ensures that \Lang{mso} atoms hold iff there exists a
witness encountered during our tree traversal, see
Equation~(\ref{red:provefin}).  Finally, Equation~(\ref{red:provefin})
is the original \Lang{mso} formula, where atoms are treated as
propositional variables.  To show the correctness of
$\mathcal{R}_{\Lang{mso}\rightarrow \Lang{qsat}}$, we rely on the
following lemma, which establishes that first-order variables get
assigned a unique element of the structure's universe.

\begin{lemma}[Correctness: Modeling of First-Order Variables]\label{lem:fo}
Given an \Lang{mso} sentence \[\phi=Q_1 S_1\dots Q_{q-1} S_{q-1} Q_{q} s_q \dots\allowbreak  Q_{\ell} s_\ell\suchthat \psi,\]
a structure~$\structure$, and a width-$k$ tree decomposition~$\mathcal{T}=(T,\bag)$.
Suppose  $\phi'=\mathcal{R}_{\Lang{mso}\rightarrow \Lang{qsat}}(\varphi, \structure, \mathcal{T})$
and let $i\in\{q,\ldots,\ell\}$.
Then:
\begin{enumerate}[(i)]
\item for any $\alpha\in 2^{\lits(\var(\varphi'))}$ and all $u,v\in
  U(\structure)$ with $u\neq v$ and ${(s_i)}_u, {(s_i)}_v\in \alpha$
  we have that $\phi'|\alpha$ is invalid iff $Q_i{=}\exists$;
\item for any $\alpha\in 2^{\lits(\var(\varphi'))}$ with
  $\alpha\cap \{{(s_i)}_u\mid u\in U(\structure)\}=\emptyset$ we have that $\phi'|\alpha$ is invalid iff $Q_i{=}\exists$.
\end{enumerate}
\end{lemma}

We prove the two items individually:
\begin{claim}
  Item~(i) holds.
\end{claim}
\begin{proof}
  By definition of TDs, there are nodes~$t_1,t_2$ of~$T$ with:
  \(
  u\in\bag(t_1)\setminus\bag(\parent(t_1))
  \)
  and
  \(
  v\in\bag(t_2)\setminus\bag(\parent(t_2)).
  \)

  \textbf{Case~$Q_i=\exists$:}
  Suppose towards a contradiction that $\phi'|\alpha$ was valid.
  Then, if~$t_1=t_2$, Equation~(\ref{redt:exclude}) is not satisfied in~$\phi'|\alpha$ for $x=s_i$ and $u'=v$. 
  Otherwise, by Equation~(\ref{redt:card}) and since $T$ is connected,
  for every node~$t^*$ on the path between~$t_1$ and~$t_2$ in~$T$, we require~$c_{\leq t^*}\in \alpha$. 
  But then either Equation~(\ref{redt:choose}) or (\ref{redt:choosechld})  for~$t_1$ or~$t_2$ is not satisfied by~$\alpha$.

  \textbf{Case~$Q_i=\forall$:}
  In contrast to the case above, we get that at least one of the negations of Equations~(\ref{redt:card})--(\ref{redt:choosechld}), which are \Lang{dnf} terms, are satisfied by~$\alpha$.
\end{proof}
\begin{claim}
  Item~(ii) holds.
\end{claim}
\begin{proof}
  We again proceed by case distinction depending on the quantifier~$Q_i$.
  
  \textbf{Case~$Q_i=\exists$:} Suppose towards a contradiction that
  $\phi'|\alpha$ was valid. However, by
  Equation~(\ref{redt:cardroot}), $c_{\leq
    \rootOf(T)}^{s_i}\in\alpha$. Consequently, by
  Equation~(\ref{redt:card}) there exists a node~$t$ in~$T$ and
  $u\in\bag(t)\setminus\bag(\parent(t))$ such that~${(s_i)}_u\in
  \alpha$.  This contradicts the assumption.

  \textbf{Case~$Q_i=\forall$:} Similarly to above and since $\phi'|\alpha$ is invalid if $Q_i=\exists$, we follow that $\phi'|\alpha$ is valid in case of~$Q_i=\forall$. 
\end{proof}     

Below we show that Figure~\ref{figt:redp} takes care of
the equivalence between \Lang{mso} atom~$\iota$ and the propositional variable for such an atom. 
In order to do so, we require the following definition.
\begin{definition}
Let~$\alpha$ be an assignment of an \Lang{mso} sentence~\[\phi=Q_1 S_1\dots Q_{q-1}\allowbreak S_{q-1} Q_{q} s_q \dots\allowbreak Q_{\ell} s_\ell\suchthat \psi\] and~$\structure$ be a structure.
Then,
we define the compatible assignment of~$\alpha$ by
\begin{align*}
  \alpha' =\, &\{\,
  X_u, \neg X_v \mid X\in\{S_1,\dots,Q_{q-1} \},
  u\in\alpha(X), v\in U(\structure)\setminus \alpha(X)\,\}\\
  \cup\, &\{\,
  x_u,
  \neg x_v \mid x\in \{s_q,\dots, s_\ell\},  u=\alpha(x), v\in
  U(\structure)\setminus\{\alpha(x)\}\,\}.
\end{align*}
\end{definition}

\begin{lemma}[Correctness: Modeling of the MSO Atoms]\label{lem:mso}
Suppose a given \Lang{mso} sentence $\phi=Q_1 S_1\dots
Q_{q-1}\allowbreak S_{q-1} Q_{q} s_q \dots\allowbreak Q_{\ell}
s_\ell\suchthat \psi$, a structure~$\structure$, and tree
decomposition~$\mathcal{T}=(T,\bag)$ of~$\structure$.
Assume $\phi'=\mathcal{R}_{\Lang{mso}\rightarrow\Lang{qsat}}(\varphi,
\mathcal{S}, \mathcal{T})$, and let $\iota\in\edges(\varphi)$ and
$\alpha$ be any assignment of~$\phi$.
Then, $\alpha$ witnesses $\structure\models\iota$ iff for the corresponding assignment~$\alpha'$ of~$\alpha$, $(\phi'\wedge \iota)|\alpha'$ is valid.
\end{lemma}
\begin{proof}
The construction of Figure~\ref{figt:redp} evaluates~$\iota$ along $\mathcal{T}$. This is correct due to connectedness of
$\mathcal{T}$ and since the elements of every relation
in~$\mathcal{S}$ appear in at least one bag.
\end{proof}

The previous two lemmas allow us to establish correctness.
In order to apply an induction, we assume for an \Lang{mso} sentence a vector~$\vec S$ of second- and first-order variables,
which will be free (unbounded) variables that do not appear in the
scope of a corresponding quantifier of the formula. That is, we
``move'' the outer most quantifers out of the sentence and make them
free variables (resulting in an \Lang{mso} formula).
For a vector~$\vec S$, by~$\vec S_i$ and $|\vec S|$ we refer to its $i$-th component and size, respectively.

\begin{lemma}[Correctness]\label{thm:corr}
Given a sequence~$\vec S$ of second-order and first-order variables,
an \Lang{mso} formula~\[\phi(\vec S)=Q_1 S_1\dots Q_{q-1} S_{q-1} Q_{q}
s_q \dots\allowbreak Q_{\ell} s_\ell\suchthat \psi\] over free
variables~$\vec S$, and a structure~$\structure$ with a corresponding
width-$k$ tree decomposition ~$\mathcal{T}=(T,\bag)$.

Suppose that $\phi'=\mathcal{R}_{\Lang{mso}\rightarrow\Lang{qsat}}(\varphi,\structure, \allowbreak \mathcal{T})$ 
is the \Lang{qbf} obtained by reduction $\mathcal{R}_{\Lang{mso}\rightarrow\Lang{qsat}}$.
Then, for any instantiation~$\vec\kappa$ of~$\vec S$ over universe~$U(\mathcal{S})$, we have $\structure\models \phi(\vec\kappa)$ iff the following is valid:
\[
\psi'=\phi'\wedge \bigwedge_{i=1}^{|\vec S|}(\bigwedge_{e\in\vec\kappa_i}{({\vec S}_i)}_e \wedge \hspace{-1em}\bigwedge_{e\in U(\structure)\setminus \vec\kappa_i}\hspace{-1em}\neg {({\vec S}_i)}_e).
\]
\end{lemma}
\begin{proof}
We proceed by induction on the number of quantifiers~$\ell$.  Base
case~$\ell=1$: Suppose that $\structure\models \phi(\vec\kappa)$.  We
sketch the result for case~$Q_\ell=\exists$, as the case
for~$Q_\ell=\forall$ works similarly. Then, since~$\ell=1$, we focus
on the case where $Q_\ell$ is a first-order quantifier, as for
second-order quantifiers~$Q_\ell=Q_1$, the formula trivially holds due
to the absence of first-order variables.

\begin{description}
\item[$\Longrightarrow$] By the semantics of \Lang{mso} model
  checking, there exists an element $v\in U(\structure)$ such that
  $\structure\models \phi^*(\vec\kappa \sqcup (s_\ell \mapsto v))$,
  where~$\phi^*$ is obtained from~$\phi$ by removing the outer-most
  quantifier~$Q_\ell s_\ell$. We construct the following assignment:
  \begin{align*}
    \alpha' =\, &\{\,
    (\vec S_i)_e, \neg (\vec S_i)_f \mid 1\leq i\leq |\vec S|, e\in\vec\kappa_i, f\in U(\structure)\setminus\kappa_i\,\}\\
    \cup\,
    &\{\,{s_\ell}_v\,\} \cup \{\,\neg{s_\ell}_u\mid u\in U(\structure), u\neq v\,\}\\
    \cup\,
    &\{\, c_{\leq t}^{s_\ell}, c_{\leq t'}^{s_\ell}, \neg c_{\leq
      t''}^{s_\ell}  \mid u\in \bag(t), t'\text{ ancestor of }t,
    t''\text{
      descendant of }t, u\notin\bag(t'') \,\}.
  \end{align*}
  Finally, we construct~$\alpha=\alpha' \cup \beta$, where~$\beta$
  contains~$p_t^\iota$ for any~$\iota\in\edges(\varphi) $ iff the formula on the right-hand side of
  Equations~(\ref{red:proveeq})--(\ref{red:provee}) hold for a
  node~$t$ of~$T$. Further, in the same manner, $\beta$ contains
  $p_{\leq t}^\iota$ and $\iota$, as required by
  Equations~(\ref{red:provedown}) and~(\ref{red:provefin}),
  respectively.  It is easy to see that~$\alpha'$ then indeed
  witnesses the validity of~$\phi'$.
  
\item[$\Longleftarrow$] There exists an assignment~$\alpha'$
  witnessing the validity of the given QBF~$\psi'$. By
  Lemma~\ref{lem:fo}, there is a unique element~$v\in U(\structure)$
  such that~${s_\ell}_{v}\in\alpha'$.  Further, since by
  Lemma~\ref{lem:mso} $\alpha'$ correctly assigns the propositional
  variable of every \Lang{mso} atom, we conclude that the first-order
  assignment~$s_\ell \mapsto v$ indeed witnesses $\structure \models
  \exists s_\ell\suchthat\varphi(\vec\kappa)$.
\end{description}
For the induction step~$\ell{-}1\rightarrow\ell$ suppose the result holds
for~$\ell-1$. We modify the \Lang{mso} formula by prepending one
of the quantifier types $\exists S_0, \forall S_0, \exists s_0,$
or $\forall s_0$. We sketch the result for quantifier
type~$\exists S_0$, as the other cases work similarly.
\begin{description}
\item[$\Longrightarrow$] By the semantics of \Lang{mso} model
  checking, we know that then there exists~$V\subseteq
  U(\structure)$ such that~$\structure\models\varphi^*(\kappa
  \sqcup(S_0\mapsto V))$, where $\varphi^*$ is obtained
  from~$\varphi$ by removing the outer-most quantifier~$\exists
  S_0$.  By the induction hypothesis, the result holds assuming that
  $S_0$ was a free (unbounded) second-order variable.  However,
  since the existence of~$V$ actually follows the semantics of an
  existential quantifier depending on free variables~$\vec S$, the
  result follows.

\item[$\Longleftarrow$] Suppose that $\psi'$ is valid.  Then, there
  exists an assignment~$\alpha'$ over variables of the outer-most
  existential quantifier, such that~$\psi'|\alpha'$ is valid.  Then,
  the QBF obtained from~$\psi'$ by removing the outer-most
  quantifier block is valid.  By induction hypothesis, we have that
  then also the \Lang{mso} obtained from~$\varphi(\vec S)$ by
  removing the outer-most quantifier is valid.  Since~$\psi'$ is
  valid, there also has to exist an assignment~$\alpha''$
  of~$\alpha'$ that witnesses the validity of~$\psi'$.  Then, by
  Lemma~\ref{lem:mso}, any such assignment $\alpha''$ correctly
  assigns the propositional variable of every occurring \Lang{mso}
  atom (independent of the remaining quantifiers).  Consequently,
  the second-order assignment~$S_0 \mapsto \{e\mid {S_0}_e\in
  \alpha'\}$ indeed witnesses $\structure \models \exists
  S_0\suchthat\varphi(\vec\kappa)$.\qedhere
\end{description}
\end{proof}

Observe that for formulas whose inner-most quantifier is existential,
i.\,e., $Q_\ell=\exists$, we immediately get the same quantifier
alternations, which by the lemmas of this section allows us to prove
Lemma~\ref{lemma:mctosat}.  However, if the inner-most quantifier were
universal, i.\,e., $Q_\ell=\forall$, Figure~\ref{figt:redp} caused an
additional quantifier block.  To avoid this, we need to
slightly modify the constructed formula and apply the normalization of
Lemma~\ref{lemma:cnf-to-dnf}:

\begin{lemma}[Quantifier Alternation Preservation]\label{thm:quantifier-preservation}
  Given an \Lang{mso} formula
  \[
  \phi = Q_1 S_1\dots Q_{q-1}\allowbreak S_{q-1} Q_{q} s_q \dots\allowbreak  Q_{\ell} s_\ell\suchthat \psi,
  \]
a structure~$\structure$ and a width-$k$ tree decomposition~$\mathcal{T}$. 
Suppose $\phi'=\mathcal{R}_{\Lang{mso}\rightarrow\Lang{qsat}}(\varphi,\structure, \allowbreak \mathcal{T})$ 
is the \Lang{qbf} obtained by the reduction.
Then there is a \Lang{qbf} $\phi''$ such that~$\qa(\phi'')=\qa(\phi)$ that is equivalent to~$\phi'$, i.\,e., for any assignment~$\alpha\sqsubseteq\var(\phi')$ we have $\phi'|\alpha$ is valid iff $\phi''|\alpha$ is valid.
\end{lemma}
\begin{proof}
  The result is trivial for~$Q_\ell=\exists$, as then~$\phi''=\phi'$.
  Assume therefore~$Q_\ell=\forall$ and let $\phi'=Q_1 S_1'\dots
  Q_{\ell} S'_\ell \exists E' \suchthat \psi$.  We partition~$\psi$
  into~$\psi_{\text{\text{def}}}$, which contains the definitions of
  every auxiliary variable in~$E'$, namely Equations~(\ref{redt:card})
  and (\ref{red:proveeq})--(\ref{red:provefin}), and the
  remainder~$\psi_{\Lang{cnf}}=\psi\setminus\psi_{\text{def}}$
  (viewing formulas as sets of clauses as defined in the
  preliminaries).

  We will modify~$\phi'$ to obtain~$\phi''=Q_1
  S_1'\dots Q_{\ell} S'_\ell \forall E' \forall F' \suchthat \neg
  \psi_{\text{def}} \vee \psi_{\Lang{dnf}}$, where~$\psi_{\Lang{dnf}}$
  is the \Lang{dnf} formula constructed from $\psi_{\Lang{cnf}}$ using
  Lemma~\ref{lemma:cnf-to-dnf} and~$F'=\var(\neg \psi_{\text{def}}
  \vee \psi_{\Lang{dnf}})\setminus \var(\psi)$.  Note that the whole
  formula $\neg \psi_{\text{def}} \vee \psi_{\Lang{dnf}}$ is in
  \Lang{dnf} as required.  While we know from
  Lemma~\ref{lemma:cnf-to-dnf} that the transition from~$\psi_{\Lang{cnf}}$
  to $\forall F' \suchthat \psi_{\Lang{dnf}}$ is correct, it still
  remains to argue why we can transform $\exists E'\suchthat
  \psi_{\text{def}}$ into $\forall E'\suchthat \neg\psi_{\text{def}}$
  in formula~$\phi''$.  However, observe that by
  Equations~(\ref{redt:card}) and
  (\ref{red:proveeq})--(\ref{red:provefin}), for any
  assignment~$\alpha\sqsubseteq \var(\psi)\setminus E'$ there is
  exactly one unique assignment~$\beta\sqsubseteq E'$ such that
  $\psi|(\alpha\cup\beta)=\emptyset$.  Equations~(\ref{redt:card}) and
  (\ref{red:proveeq})--(\ref{red:provefin}) are constructed for every
  node~$t$ in~$T$, i.\,e., these equations can be ordered such that we
  can apply those from leaf nodes towards the root. Thereby we apply
  equivalences and inductively construct assignment~$\beta$ from
  $\alpha$ and already assigned variables in~$\beta$.
  Since this assignment~$\beta$ is unique, we might as well
  universally quantify the variables contained in~$\beta$ and ensure
  that, if we do not get this precise assignment~$\beta$, the required
  formula remains satisfied.
  This is precisely the construction of~$\phi''$, where by~$\neg
  \psi_{\text{def}}$ we either disregard assignments~$\beta'\sqsubseteq E'$
  such that~$\beta'\neq\beta$, or we use~$\beta$ and then require to
  satisfy the whole \Lang{cnf} formula~$\psi_{\Lang{cnf}}$, which by
  Lemma~\ref{lemma:cnf-to-dnf} is ensured using $\psi_{\Lang{dnf}}$
  via fresh auxiliary variables~$F'$.
\end{proof}

We are left with the task of binding the treewidth of the produced
formula. The following lemma establishes the claimed bound of
$9|\phi|+k|\phi|$. From the proof of the lemma, it is also immediately
apparent how to transform the given tree decomposition into a tree
decomposition of $\psi$.

\begin{lemma}\label{lemma:qbf-twbound}
  The \Lang{qbf} $\psi$ constructed by Lemma~\ref{lemma:mctosat} has treewidth at most~$9|\phi|+k|\phi|$.
\end{lemma}
\begin{proof}
  We prove the following more general result:
  \begin{claim}
    Given a sentence $\phi=Q_1 S_1\dots Q_{q-1} S_{q-1} Q_{q} s_q\allowbreak \dots  Q_{\ell} s_\ell\suchthat \psi$,
    a structure~$\structure$, and a width-$k$ tree
    decomposition~$\mathcal{T}=(T,\bag)$. Then, $\mathcal{R}_{\Lang{mso}\rightarrow\Lang{qsat}}(\varphi,\structure, \mathcal{T})$
    constructs a \Lang{qbf}~$\phi'=Q_1 S'_1\dots Q_{\ell} S'_\ell\suchthat \psi'$ such that the treewidth of~$\psi'$
    is bounded by~$6|\edges(\psi)
    |+3\ell+\ell\cdot k$.
  \end{claim}
  \begin{proof}
    We construct a tree decomposition~$\mathcal{T}'=(T,\bag')$ of the
    primal graph of the~$\psi$, where~$\bag'$ is defined as follows.
    For every node~$t$ in~$T$, we define:
    \begin{align*}
      \bag'(t) =\, &\{\,
    c_{\leq t}^x,  c_{\leq t'}^x, x_u \mid u\in \bag(t), x\in \{s_q,
    \ldots, s_\ell\}, t'\in\children(t)
    \,\}\\   
    &\{\,
    p_t^{X(x)}, p_{\leq t}^{X(x)}, p_t^{E(x,y)}, p_{\leq t}^{E(x,y)},
    p_{t'}^{X(x)}, p_{\leq t'}^{X(x)}, p_{t'}^{E(x,y)},
    p_{\leq t'}^{E(x,y)}, E(x,\allowbreak y), X(x) \mid\\
    &\qquad
    x,y\in \{s_q, \ldots, s_\ell \}; X(x), E(x,y)\in \edges(\psi); t'\in\children(t)
    \,\}.
    \end{align*}
    Indeed, $\mathcal{T'}$ is connected and a well-defined tree decomposition.
    In particular, Equations~(\ref{redt:card})--(\ref{red:form}) are covered, i.\,e.,
    the variables of every clause appear in at least one node.
  \end{proof}
  Therefore, since every bag is bounded by $6|\edges(\psi)
    |+3\ell+\ell\cdot k$,  the treewidth is bounded by $6|\edges(\psi)
    |+3\ell+\ell\cdot k \leq 9|\phi|+k|\phi| = |\phi| (9+k)$.
\end{proof}

\end{document}